%% file: maglim.tex
\pdfoutput=1
\documentclass[aps,prd,amsmath,amssymb,superscriptaddress, nofootinbib, longbibliography, showkeys, reprint,preprintnumbers]{revtex4-2}

\usepackage{color}
\usepackage{aas_macros}
\usepackage{hyperref,breakurl}
\hypersetup{colorlinks=true, citecolor=blue, urlcolor=blue, linkcolor=blue}
\usepackage[usenames,dvipsnames]{xcolor}
\usepackage{dcolumn}
\usepackage{bm}
\usepackage{array}
\usepackage{graphicx}
\usepackage{graphics}
\usepackage{enumitem}
\usepackage{xspace}
\usepackage{float}

\usepackage{amsmath,amssymb,latexsym,times}

\usepackage{todonotes}
\usepackage[normalem]{ulem} 

\usepackage{microtype}

\newcommand{\beq}{\begin{equation}}
\newcommand{\eeq}{\end{equation}}

\newcommand{\code}[1]{\texttt{#1}\xspace}

\newcommand{\ngmix}{\textsc{ngmix}}

\newcommand{\dnf}{\texttt{DNF}}
\newcommand{\buzzard}{\textsc{Buzzard}\xspace}

\newcommand{\redmagic}{\textsc{redMaGiC}\xspace}
\newcommand{\maglim}{\textsc{MagLim}\xspace}
\newcommand{\metacal}{\textsc{Metacalibration}\xspace}

\newcommand{\gold}{\code{Y3\,GOLD}}

\newcommand{\photoz}{photo-$z$}

\newcommand{\LCDM}{$ \Lambda $CDM\xspace}
\newcommand{\wCDM}{$w$CDM\xspace}

\newcommand{\ode}{\Omega_{\rm de}}
\newcommand{\omb}{\Omega_{\rm b}}

\newcommand{\mpc}{h^{-1}\mathrm{Mpc}}

\newcommand{\sqdeg}{{\rm deg}^{2}}

\newcommand{\nside}{\ifmmode N_{\mathrm{side}}\else $N_{\mathrm{side}}$\fi}
\newcommand{\npix}{\ifmmode n_{\mathrm{pix}}\else $n_{\mathrm{pix}}$\fi}
\newcommand{\Npix}{\ifmmode N_{\mathrm{pix}}\else $n_{\mathrm{pix}}$\fi}
\newcommand{\lmin}{\ifmmode \ell_{\mathrm{min}}\else $\ell_{\mathrm{min}}$\fi}
\newcommand{\lmax}{\ifmmode \ell_{\mathrm{max}}\else $\ell_{\mathrm{max}}$\fi}

\newcommand{\mr}[1]{\mathrm{#1}}

\newcommand\Tstrut{\rule{0pt}{2.6ex}}         

\definecolor{wine}{RGB}{153,60,51}

\begin{document}

\title{Dark Energy Survey Year 3 results: Cosmological constraints from galaxy clustering and galaxy-galaxy lensing using the $\maglim$ lens sample }

\author{A.~Porredon}
\affiliation{Center for Cosmology and Astro-Particle Physics, The Ohio State University, Columbus, OH 43210, USA}
\affiliation{Department of Physics, The Ohio State University, Columbus, OH 43210, USA}
\affiliation{Institute of Space Sciences (ICE, CSIC),  Campus UAB, Carrer de Can Magrans, s/n,  08193 Barcelona, Spain}
\affiliation{Institut d'Estudis Espacials de Catalunya (IEEC), 08034 Barcelona, Spain}
\author{M.~Crocce}
\affiliation{Institute of Space Sciences (ICE, CSIC),  Campus UAB, Carrer de Can Magrans, s/n,  08193 Barcelona, Spain}
\affiliation{Institut d'Estudis Espacials de Catalunya (IEEC), 08034 Barcelona, Spain}
\author{J.~Elvin-Poole}
\affiliation{Center for Cosmology and Astro-Particle Physics, The Ohio State University, Columbus, OH 43210, USA}
\affiliation{Department of Physics, The Ohio State University, Columbus, OH 43210, USA}
\author{R.~Cawthon}
\affiliation{Physics Department, 2320 Chamberlin Hall, University of Wisconsin-Madison, 1150 University Avenue Madison, WI  53706-1390}
\author{G.~Giannini}
\affiliation{Institut de F\'{\i}sica d'Altes Energies (IFAE), The Barcelona Institute of Science and Technology, Campus UAB, 08193 Bellaterra (Barcelona) Spain}
\author{J.~De~Vicente}
\affiliation{Centro de Investigaciones Energ\'eticas, Medioambientales y Tecnol\'ogicas (CIEMAT), Madrid, Spain}
\author{A.~Carnero~Rosell}
\affiliation{Instituto de Astrofisica de Canarias, E-38205 La Laguna, Tenerife, Spain}
\affiliation{Laborat\'orio Interinstitucional de e-Astronomia - LIneA, Rua Gal. Jos\'e Cristino 77, Rio de Janeiro, RJ - 20921-400, Brazil}
\affiliation{Universidad de La Laguna, Dpto. Astrofísica, E-38206 La Laguna, Tenerife, Spain}
\author{I.~Ferrero}
\affiliation{Institute of Theoretical Astrophysics, University of Oslo. P.O. Box 1029 Blindern, NO-0315 Oslo, Norway}
\author{E.~Krause}
\affiliation{Department of Astronomy/Steward Observatory, University of Arizona, 933 North Cherry Avenue, Tucson, AZ 85721-0065, USA}
\author{X.~Fang}
\affiliation{Department of Astronomy/Steward Observatory, University of Arizona, 933 North Cherry Avenue, Tucson, AZ 85721-0065, USA}
% ...

\collaboration{DES Collaboration}

\noaffiliation

\date{\today}


\begin{abstract}
The cosmological information extracted from photometric surveys is most robust when multiple probes of the large scale structure of the universe are used. Two of the most sensitive probes are the clustering of galaxies and the tangential shear of background galaxy shapes produced by those foreground galaxies, so-called galaxy-galaxy lensing. Combining the measurements of these two two-point functions leads to cosmological constraints that are independent of the way galaxies trace matter (the galaxy bias factor). The optimal choice of foreground, or {\it lens}, galaxies is governed by the joint, but conflicting requirements to obtain accurate redshift information and large statistics.  We present cosmological results from the full 5000 deg$^2$ of the Dark Energy Survey first three years of observations (Y3) combining those two-point functions, using for the first time a magnitude-limited lens sample ($\maglim$) of 11 million galaxies especially selected to optimize such combination, and 100 million background shapes. We consider two cosmological models, flat \LCDM and \wCDM, and marginalized over 25 astrophysical and systematic nuisance parameters. In \LCDM we obtain for the matter density $\Omega_m = 0.320^{+0.041}_{-0.034}$  and for the clustering amplitude $S_8\equiv \sigma_8 (\Omega_m/0.3)^{0.5} = 0.778^{+0.037}_{-0.031}$, at 68\% C.L. The latter is only 1$\sigma$ smaller than the prediction in this model informed by  measurements of the cosmic microwave background by the \textit{Planck} satellite. In \wCDM we find $\Omega_m = 0.32^{+0.044}_{-0.046}$, $S_8=0.777^{+0.049}_{-0.051}$,  and dark energy equation of state $w=-1.031^{+0.218}_{-0.379}$. We find that including smaller scales while marginalizing over non-linear galaxy bias improves the constraining power in the $\Omega_m-S_8$ plane by $31\%$ and in the $\Omega_m-w$ plane by $41\%$ while yielding consistent cosmological parameters from those in the linear bias case. 
These results are combined with those from cosmic shear in a companion paper to present full DES-Y3 constraints from the three two-point functions ($3\times2$pt).
\end{abstract}

\keywords{dark energy; dark matter; cosmology: observations; cosmological parameters}

\preprint{DES-2020-0620}
\preprint{FERMILAB-PUB-21-251-AE}

\maketitle

\section{Introduction}
\label{sec:intro}

The discovery of the accelerated expansion of the universe in the 
1990s has opened one of the most enduring and widely-researched questions in
the field of cosmology: what is the nature of the physical process that powers
the acceleration? The source of this increasing expansion rate --- a new energy density component, called
dark energy --- has become a key part of the cosmic inventory, yet its
physical nature and microphysical properties are  unknown. Over the
course of about two decades since the discovery of dark energy, an impressive
variety of measurements from cosmological probes has helped to set tighter
constraints on its energy density relative to the critical density, $\ode$, and its
equation of state ratio $w=P_{\rm de}/\rho_{\rm de}$, where $P_{\rm de}$ and $\rho_{\rm de}$ are, respectively, the pressure and energy density of dark energy. These probes include
distance measurements to Type Ia supernovae
(SNIa) \cite{1998AJ....116.1009R,1999ApJ...517..565P}, cosmic microwave
background fluctuations (CMB) \cite{2003ApJS..148..175S,2018arXiv180706209P}
and the study of the large-scale structure (LSS) in our Universe.  The latter
carries a wealth of cosmological information and allows for tests of the fiducial cold-dark-matter plus dark energy cosmological model, $\Lambda$CDM
(e.g.  \cite{2006PhRvD..74l3507T,2017MNRAS.470.2617A,2019PhRvL.122q1301A,2020arXiv200708991E,2017MNRAS.464.1640S,2020JCAP...05..042I,2020JCAP...05..005D}
and references therein).

In the past few years, early results from Stage-III dark energy surveys have been released, significantly improving the quality and quantity of data and the strength
of cosmological constraints from LSS probes of dark energy. The
Stage-III surveys include the Dark Energy Survey
(DES\footnote{\url{http://www.darkenergysurvey.org/}}) \cite{DESY1_3x2,2019PhRvL.122q1301A},
the Kilo-Degree Survey
(KiDS\footnote{\url{http://kids.strw.leidenuniv.nl/}}) \cite{Joudaki2018,vanUitert2018},
Hyper Suprime-Cam Subaru Strategic Program (HSC-SSP\footnote{\url{https://hsc.mtk.nao.ac.jp/ssp/}}) \cite{Hikage2019,Hamana2020}. These
surveys have demonstrated the feasibility of ambitious photometric LSS
analyses, and featured extensive testing of theory, inclusion of a large
number of systematic parameters in the analysis, and blinding of the analyses
before the results are revealed. These photometric LSS surveys have (so far)
confirmed the \LCDM model and tightened the constraints on some of the key
cosmological parameters. On the other hand, these surveys have also begun to
reveal an apparent tension between the measurements of the parameter
$S_8\equiv \sigma_8 (\Omega_{\rm m}/0.3)^{0.5}$, the amplitude of mass fluctuations
$\sigma_8$ scaled by the square root of matter density $\Omega_{\rm m}$. This is
measured to be higher in the CMB ($S_8\simeq 0.834\pm 0.016$ \cite{2018arXiv180706209P}) than in
photometric surveys, including the Dark Energy Survey Year 1 (Y1) result,
$S_8= 0.794\pm0.028$ \cite{DESY1_3x2}.

New and better data will be key to bring these tensions into sharp focus in
order to see if they are due to new physics. The next generation of LSS
surveys that will provide high quality data include the Rubin Observatory Legacy
Survey of Space and Time
(LSST\footnote{\url{https://www.lsst.org/}}) \cite{LSST},
Euclid\footnote{\url{https://sci.esa.int/web/euclid}} \cite{Euclid}, and the
Nancy Grace Roman Space Telescope
(Roman\footnote{\url{https://roman.gsfc.nasa.gov/}}) \cite{Spergel2015}. These
upcoming surveys will map the structure in the universe over a wider and deeper range of
temporal and spatial scales (see e.g. \cite{Eifler2020A,Eifler2020B}). Two key cosmological
probes that all of these surveys will use are galaxy clustering and weak
gravitational lensing.

When selecting a sample of objects to use in a photometric survey, there is a
trade-off between selecting the \textit{largest} galaxy sample (to
minimize shot noise), and a sample with the best \textit{redshift accuracy},
which generally includes only a small subset of galaxies. The latter strategy
typically uses luminous red galaxies (LRGs), which are characterized by a
sharp break at 4000\textup{\AA} \cite{Eisenstein2001,2007MNRAS.378..852P}. LRGs have a
remarkably uniform spectral energy distribution and correlate strongly with
galaxy cluster positions. Such an approach was taken in the DES Y1
analysis \cite{DESY1_3x2}, where lens galaxies were selected using
the \redmagic\ algorithm \cite{Redmagic}, which relies on the calibration of
the red-sequence in optical galaxy clusters. The KiDS survey recently made a
similar selection of red-sequence galaxies \cite{KiDS_LRGs_2020}, and such
selection of LRGs in photometric data has also been adopted for measurements
of baryon acoustic oscillations \cite{2007MNRAS.378..852P,
2019MNRAS.482.2807C,2020arXiv200513126S}.

An alternative strategy is to select the largest galaxy sample
possible. Selecting all galaxies up to some limiting magnitude leads to a
galaxy sample that reaches a higher redshift, has a much higher number
density, but also less accurate redshifts (larger photo-$z$ errors). Such
flux-limited samples have been used in the DES Science Verification
analysis \cite{Crocce2016} and, previously, in the galaxy clustering
measurements from Canada-France-Hawaii Telescope Legacy Survey (CFHTLS)
data \cite{Coupon2012}; these two analyses had an upper apparent magnitude cut
of $i<22.5$. More recently,  \citet{Nicola2020} also selected galaxies with a limiting magnitude ($i<24.5$) from the first HSC public data release to analyze the  galaxy clustering and other properties of that sample, such as large-scale bias.
This kind of galaxy selection is simple and easily reproducible in different datasets, and
leads to a sample whose properties can be well understood. For instance,
\citet{Crocce2016} showed that the redshift evolution of the linear galaxy
bias of their sample matches the one from CHFTLS \cite{Coupon2012}, and is
also consistent with that from HSC data \cite{Nicola2020}. However, this type
of selection that selects the largest possible galaxy sample  has not yet been
used to produce constraints on cosmological parameters.

The DES collaboration recently investigated potential gains in using such a
magnitude-limited sample in simulated data in \citet{y3-2x2maglimforecast}. We
assumed synthetic DES Year 3 (Y3) data and the DES Year 1 \metacal\ sample of
source galaxies, and explored the balance between density and photometric
redshift accuracy, while marginalizing over a realistic set of cosmological
and systematic parameters.  The optimal sample, dubbed the \maglim\ sample,
 satisfies $i < 4 \, z_{\rm phot} + 18$ and has $\sim 30\%$ wider redshift
distributions but $\sim 3.5$ times more galaxies than \redmagic. We found an
improvement in cosmological parameter constraints of tens of percent (per
parameter) using \maglim\ relative to an equivalent analysis using \redmagic.
Finally, we showed that the results are robust with respect to the assumed
galaxy bias and photometric redshift uncertainties.

In this paper, we show cosmological results from DES Y3 data
using the \maglim\ sample. We specifically consider galaxy clustering and
galaxy-galaxy lensing, that is, the auto-correlations of \maglim\ galaxies' positions and
their cross-correlation with cosmic shear (2$\times$2pt). 
This analysis is complemented by two other papers that combine these two two-point functions from DES Y3 data: an equivalent analysis using the $\redmagic$ \cite{y3-galaxyclustering} lens sample \cite{y3-2x2ptbiasmodelling}, and a study of the impact of magnification on the 2$\times$2pt cosmological constraints using both $\maglim$ and $\redmagic$ lens samples \cite{y3-2x2ptmagnification}.
In addition, the results presented in this work are combined with the cosmological analysis of cosmic shear \cite{y3-cosmicshear1,y3-cosmicshear2} in  \cite{y3-3x2ptkp} to obtain the final DES-Y3 $3\times2$pt constraints.

The paper is organized as follows. Sec.~\ref{sec:DESY3data} introduces the
data, the mask and the data vector measurements. Sec.~\ \ref{sec:photoz_calib} presents
different estimations of the photometric redshift distribution of the \maglim sample.
Sec.~\ref{sec:sim} describes the simulations used to
test the methodology and pipelines. 
Sec.~\ref{sec:method} presents the methodology. The validation of the
methodology on the simulations and theory data vectors is presented in Sec.~\ref{sec:valid}. Our main
results are presented in Sec.~\ref{sec:cosmology-constrains}, along with a
discussion of some changes made post-unblinding and robustness tests. Conclusions are presented in Sec.~\ref{sec:conclusions}.

\section{Data}
\label{sec:DESY3data}

\subsection{DES Y3}
DES is an imaging survey that has observed $\sim5000$ $\sqdeg$ of the southern sky using the Dark Energy Camera \cite{Flaugher2015} on the 4 m Blanco telescope at the Cerro Tololo Inter-American Observatory (CTIO) in Chile. DES completed observations in January 2019, after 6 years of operations in which it collected information from more than 500 million galaxies in five optical filters, $grizY$, covering the wavelength range from $\sim400$ nm to $\sim1060$ nm \cite{DESDR2}.

In this work we use data from the first three years of observations (Y3), which were taken from August 2013 to February 2016. The core dataset used in Y3 cosmological analyses, the $\gold$ catalog, is largely based on the  coadded object catalog that was released publicly as the DES Data Release 1 (DR1)\footnote{Available at \href{https://des.ncsa.illinois.edu/releases/dr1}{https://des.ncsa.illinois.edu/releases/dr1} } \cite{DESDR1}, and includes additional enhancements and data products with respect to DR1, as described extensively in \cite{y3-gold}.  The \gold catalog includes nearly 390 million objects with depth reaching S/N $\sim10$ up to limiting magnitudes of $g=24.3$, $r=24.0$, $i=23.3$, $z=22.6$, and $Y=21.4$. Objects are detected using \code{SourceExtractor} from the $r$+$i$+$z$ coadd images (see \cite{Morganson2018} for further details). The morphology and flux of the objects is determined through the multiobject fitting pipeline (MOF), and its variant single-object fitting (SOF), which simplifies the fitting process with negligible impact in performance \cite{y1-gold}. 

The SOF photometry is used to generate the photometric redshift (\photoz) estimates from different codes: \code{BPZ} \cite{Benitez:2000}, \code{ANNz2} \cite{annz2}, and \dnf \cite{DNF2016}. In this work, we rely on SOF magnitudes and \dnf\ \photoz$ $ estimates for the \maglim$ $ sample selection, which we describe below.
For the source galaxies, we use $\metacal$ photometry \cite{metacal} instead of SOF.  This photometry is measured similarly to the SOF and MOF pipelines but, while the latter use $\ngmix$ \cite{ngmix} in order to reconstruct the point spread function (PSF), $\metacal$ uses a simplified Gaussian model for the PSF.

The total area of the \gold catalog footprint comprises 4946 $\sqdeg$. For the cosmology analyses presented here we apply a masking that we describe in detail in Sec~\ref{sec:mask}, resulting in a final area of about 4143.17 $\sqdeg$. 

In the following, we describe the selection of our lens and source samples.

\subsection{Lens samples}

\begin{figure}
	\begin{center}
		\includegraphics[width=\linewidth]{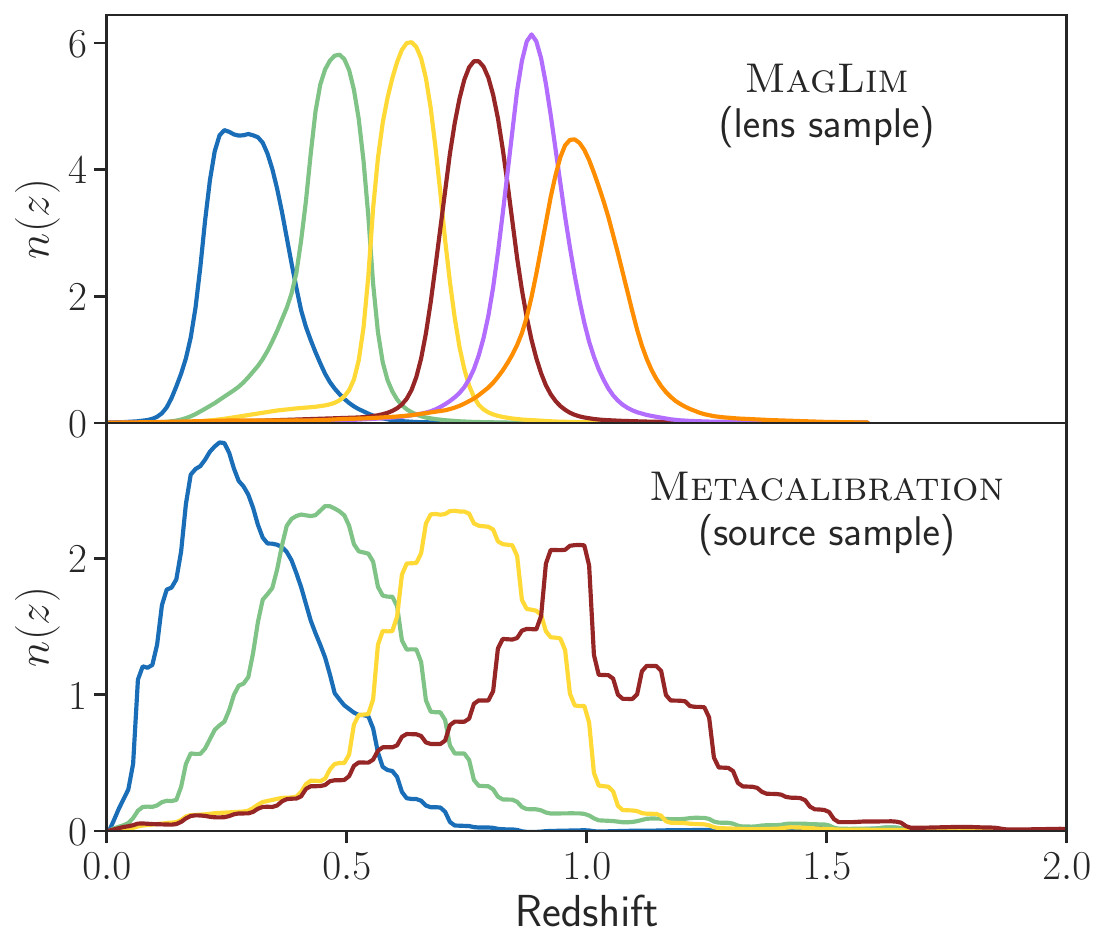}
		\caption{Redshift distributions of the $\maglim$ lens sample and the $\metacal$ source sample. The distributions are normalized to 1, i.e. we impose $\int n(z) \mathrm{d}z = 1$. }
		\label{fig: lens source nzs}
	\end{center}
\end{figure} 

In what follows, we describe the two lens samples used throughout this work, focusing on the \maglim\ sample. Both these samples present correlations of their galaxy number density with various observational properties of the survey, which themselves are correlated too. This imprints a non-trivial angular selection function for these galaxies which translates into biases in the clustering signal if not accounted for.  This is a common feature of galaxy surveys, in particular imaging surveys, and different strategies have been proposed in the literature to mitigate this contamination (e.g. see \cite{2020arXiv200714499W} for a recent review). We correct this effect by applying weights to each galaxy corresponding to the inverse of the estimated angular selection function. The computation and validation of these weights, for both $\maglim$ and $\redmagic$, is described in \citet{y3-galaxyclustering}.

\begin{table}
\centering
\caption{Summary description of the samples used in this work. $N_\text{gal}$ is the number of galaxies in each redshift bin, $n_\text{gal}$ is the effective number density (including the weights for each galaxy) in units of gal/arcmin$^{2}$, ``bias" refers to the 68\% C.L. constraints on the linear galaxy bias from the 3$\times$2pt \LCDM cosmology result described in \cite{y3-3x2ptkp}, $C$ is the magnification coefficient as measured in \cite{y3-2x2ptmagnification} and defined in Sec.~\ref{sec:model}, and $\sigma_\epsilon$ is the weighted standard deviation of the ellipticity for a single component as computed in \cite{y3-cosmicshear1}.}
\label{tab:samples}
\vspace{5mm}
Lens sample 1 : \maglim \\
\vspace{2mm}
\renewcommand{\arraystretch}{1.5}
\begin{tabular}{ccccc}
\hline
\hline		
\text{Redshift bin} &   \text{$N_\text{gal}$} &  \text{$n_\text{gal}$} & $\text{bias}$ & \text{$C$} \\
\hline	
1\,:\,$0.20 < z_{\rm ph} < 0.40$ \, & \, 2 236 473 \, & \, 0.150 \, & \, $1.40^{+0.10}_{-0.09}$ \, & \, 0.43 \\
2\,:\,$0.40 < z_{\rm ph} < 0.55$ \, & \, 1 599 500 \, & \, 0.107 \, & \, $1.60^{+0.13}_{-0.10}$ \, & \, 0.30 \\
3\,:\,$0.55 < z_{\rm ph} < 0.70$ \, & \, 1 627 413 \, & \, 0.109 \, & \, $1.82^{+0.13}_{-0.10}$ \, & \, 1.75 \\
4\,:\,$0.70 < z_{\rm ph} < 0.85$ \, & \, 2 175 184 \, & \, 0.146 \, & \, $1.70^{+0.12}_{-0.09}$ \, & \, 1.94 \\
5\,:\,$0.85 < z_{\rm ph} < 0.95$ \, & \, 1 583 686 \, & \, 0.106 \, & \, $1.91^{+0.14}_{-0.10}$ \, & \, 1.56 \\
6\,:\,$0.95 < z_{\rm ph} < 1.05$ \, & \, 1 494 250 \, & \, 0.100 \, & \, $1.73^{+0.14}_{-0.10}$ \, & \, 2.96 \\
\hline	
\end{tabular}
\\
\vspace{5mm}
Lens sample 2 : \redmagic \\
\vspace{2mm}
\begin{tabular}{ccccc} 
	\hline
	\hline	
$\text{Redshift bin}$  &  $N_\text{gal}$ &  \text{$n_\text{gal}$} & $\text{bias}$ & \text{$C$} \\ 
\hline
1\,:\,$0.15 < z_{\rm ph} < 0.35$ \, & \, 330 243 \, & \, 0.022 \, & \, $1.74^{+0.10}_{-0.13}$ \, & \, 0.63  \\
2\,:\,$0.35 < z_{\rm ph} < 0.50$ \, & \, 571 551 \, & \, 0.038 \, & \, $1.82^{+0.11}_{-0.11}$ \, & \, -3.04 \\
3\,:\,$0.50 < z_{\rm ph} < 0.65$ \, & \, 872 611 \, & \, 0.058 \, & \, $1.92^{+0.11}_{-0.12}$ \, & \, -1.33 \\
4\,:\,$0.65 < z_{\rm ph} < 0.80$ \, & \, 442 302 \, & \, 0.029 \, & \, $2.15^{+0.11}_{-0.13}$\, & \, 2.50 \\
5\,:\,$0.80 < z_{\rm ph} < 0.90$ \, & \, 377 329 \, & \, 0.025 \, & \, $2.32^{+0.13}_{-0.14}$ \, & \, 1.93 \\
\hline
\end{tabular}
\\
\vspace{5mm}
Source sample : \metacal \\
\vspace{2mm}
\begin{tabular}{ccccc}
	\hline
	\hline
\text{Redshift bin} &   \text{$N_\text{gal}$} &  \text{$n_\text{gal}$ } & \text{$\sigma_\epsilon$} & \text{$C$} \\
\hline
1 \, & \, 24 941 833 \, & \, 1.476 \, & \, 0.243 \, & \, -1.32 \\
2 \, & \, 25 281 777 \, & \, 1.479 \, & \, 0.262 \, & \, -0.62 \\
3 \, & \, 24 892 990 \, & \, 1.484 \, & \, 0.259 \, & \, -0.02 \\
4 \, & \, 25 092 344 \, & \, 1.461 \, & \, 0.301 \, & \, 0.92 \\
\hline
\end{tabular}
\renewcommand{\arraystretch}{1.0}
\end{table}

\subsubsection{\maglim}

The main lens sample considered in this work,  \maglim,  is defined with a magnitude cut in the $i$-band that depends linearly on the photometric redshift $z_{\rm phot}$, $i < 4z_{\rm phot} + 18$. This selection is the result of the optimization carried out in \citet{
y3-2x2maglimforecast} in terms of its $2\times2$pt cosmological constraints. Additionally, we apply a lower magnitude cut, $i>17.5$, to remove stellar contamination from binary stars and other bright objects. We split the sample in 6 tomographic bins from $z=0.2$ to $z=1.05$, with bin edges $[0.20, 0.40, 0.55, 0.70, 0.85, 0.95, 1.05]$.  We note that the edges have been slightly modified with respect to \cite{
y3-2x2maglimforecast} in order to improve the photometric redshift calibration\footnote{With these new bin edges we avoid having a double-peaked redshift distribution in the second tomographic bin.}. The number of galaxies in each tomographic bin and other properties of the sample are shown in Table~\ref{tab:samples}. In total, $\maglim$ amounts to 10.7 million galaxies in the redshift range considered. The abundance of the sample as a function of redshift is shown in Fig.~1 from Ref.~\cite{y3-2x2maglimforecast}. The number of galaxies remains approximately constant with redshift, with a slightly increasing trend in $0.6 < z < 0.9$. We would expect more of a monotonically decreasing trend if the sample had a flat magnitude limit (e.g. $i< 22.2$). However, the  \maglim selection depends linearly on the photometric redshift, which allows including more galaxies at higher redshift.
We refer the reader to \cite{
y3-2x2maglimforecast} for more details about the optimization of this sample and its comparison with \redmagic$ $ and other flux-limited samples. See also Sec.~\ref{sec:photoz_calib} and Appendix~\ref{sec: Comparison of DNF with VIPERS} for further information on the photometric redshift calibration and validation.

\subsubsection{\redmagic}

The other lens sample used in the DES Y3 analysis is selected with the \redmagic algorithm~\cite{y3-galaxyclustering,y3-2x2ptbiasmodelling,y3-2x2ptmagnification}. \redmagic selects Luminous Red Galaxies (LRGs) according to the magnitude-color-redshift relation of red sequence galaxy clusters, calibrated using an overlapping spectroscopic sample. 
This sample is defined by an input threshold luminosity $L_{\rm min}$ and constant comoving density. The full \redmagic algorithm is described in \cite{Redmagic}.

There are 2.6 million galaxies in the Y3 \redmagic sample, which are placed in five tomographic bins, based on the \redmagic redshift point estimate quantity ZREDMAGIC. The bin edges used are $z=[0.15, 0.35, 0.50, 0.65, 0.80, 0.90]$. 
The redshift distributions are computed by stacking samples from the redshift PDF of each individual \redmagic galaxy, allowing for the non-Gaussianity of the PDF. From the variance of these samples we find an average individual redshift uncertainty of $\sigma_z/(1+z)=0.0126$ in the redshift range used.

\subsection{Source sample}
\label{subsec:source sample}
The source sample that is used for cross-correlation with the
foreground lens samples consists of 100,204,026 galaxies with shapes
measured in the $riz$ bands {y3-shapecatalog}. The source galaxies cover the same effective area as the foreground lens tracers (after masking described below, 4143.17 $\sqdeg$),
have a weighted source number density of $n_{\rm eff}=5.59$ gal/arcmin$^2$
and shape noise $\sigma_e=0.261$ per ellipticity component.

The source shapes are measured using the \textsc{metacalibration} method \citep{Huff_Mandelbaum_2017,metacal}, which measures the response of a given shear estimator to a small applied shear.  The implementation closely follows that of the previous Y1 source shape catalog \cite{DESY1_shapes}. For each galaxy, the point-spread function is deconvolved before the artificial shear is applied, and then the image is reconvolved with a symmetrized version of the PSF.  Here, as in \cite{DESY1_shapes}, the ellipticities are calculated from single Gaussians using the \textsc{ngmix} software\footnote{\url{https://github.
com/esheldon/ngmix}}.  The PSF models used in the aforementioned
deconvolution step have been measured with the PSFs In the Full FOV
(\textsc{piff}) software \cite{y3-piff}. \citet*{y3-shapecatalog} provides a full
account of the catalog creation and a set of validation tests, including checks for B modes and correlations between shape measurements and a number of galaxy and survey properties. An
accompanying paper \cite{y3-imagesims} calibrates the shear
measurement pipeline on a suite of realistic image simulations.  The relationship between an input shear, $\gamma$ and measured shape, $\epsilon^{\rm obs}$, is given by:
\begin{equation}
\label{eqn:shearbias}
    \epsilon^{\rm obs} = (1+m)(\epsilon^{\rm int} + \gamma) + c\, .
\end{equation}
\citet{y3-imagesims} determines the multiplicative bias, $m$, and the additive bias, $c$, using our full object detection and shape measurement pipeline.  $\epsilon^{\rm int}$ is the intrinsic galaxy shape, part of which is random with mean zero and variance $\sigma_e^2$ and the other part of which is due to {\it intrinsic alignment}, discussed in Sec.~\ref{sec:valid}.  Note that ellipticity and shear have two components, so Eq.~\eqref{eqn:shearbias} is often written with appropriate indices, suppressed here.

The source sample is sub-divided into four tomographic bins, with corresponding redshift distributions and uncertainties derived in \citet*{y3-sompz} using the Self-Organizing Map Photometric Redshift (SOMPZ) method. The cross-correlation redshift (WZ) approach provides further calibration, as described in \citet*{y3-sourcewz}. The `source sample' section of Table~\ref{tab:samples} provides the number of galaxies, densities, and shape noise for the source galaxies separated into the SOMPZ-defined redshift bins (more details in Table~I from \cite{y3-cosmicshear1}) .

\subsection{Mask}
\label{sec:mask}

As mentioned previously the area of the \gold catalog footprint spans 4946 $\sqdeg$. However additional masking is imposed to remove 
regions with either astrophysical foregrounds (bright stars or nearby galaxies) or with recognised data processing issues (`bad regions'). This is achieved by a set of flags that we describe below, leading to a reduction of area by 659.68 $\sqdeg$ \cite{y3-gold}. This mask is defined on a pixelated healpix map \cite{healpix} of resolution $4096$. From that map we remove pixels with fractional coverage less than 80$\%$. Lastly, we ensure that both samples used for clustering have homogeneous depth across the footprint in all redshift bins by removing shallow and incomplete regions, using the corresponding limiting depth maps (or the quantity ZMAX in the case of \redmagic). In all, the \gold catalog quantities \cite{y3-gold} we select on to define the final mask are summarised by,

\begin{itemize}
    \item footprint $>=$ 1
    \item foreground $==$ 0
    \item badregions $<=$ 1
    \item fracdet $>$ 0.8
    \item depth $i$-band $>=$ 22.2
    \item ZMAX$_{\rm highdens} > $ 0.65
    \item ZMAX$_{\rm highlum} > $ 0.90 
\end{itemize}
where depth $i$-band corresponds to SOF photometry (as used in \maglim) and the conditions on ZMAX are inherited from the \redmagic redshift span. For simplicity, we apply the same mask for all our samples, resulting in a final effective area of 4143.17 $\sqdeg$. 

\subsection{Data-vector measurements}
\label{subsec: DVs}

We are extracting cosmological  information using the combination of two two-point correlation functions: 
(1) the auto-correlation of angular positions of lens galaxies (a.k.a. {\it galaxy clustering}) and (2) the cross correlation of lens galaxy positions and source galaxy shapes (a.k.a. {\it galaxy galaxy-lensing}). These angular correlation functions are computed after the galaxies have been separated into tomographic bins, as presented in Table \ref{tab:samples}. 

\textit{Galaxy Clustering}: The two-point function between galaxy positions in redshift bins $i$ and $j$, $w^{ij}(\theta)$, describes the excess (over random) number of galaxies separated by an angular distance $\theta$. Our fiducial result uses only the auto correlation of galaxies in the same bin ($i=j$).
This correlation is measured in 20 logarithmic angular bins between 2.5 and 250 arcmin. Some of these bins are removed after imposing scale cuts, see Sec.~\ref{sec:scale_cuts}, leaving a total data vector size of 69 elements for \maglim and 54 for \redmagic (only auto-correlations on linear scales). The validation and robustness of the clustering signal measurement for both \maglim and \redmagic is presented in detail in \citet{y3-galaxyclustering}. 

\textit{Galaxy--Galaxy Lensing}: The two-point function between lens galaxy positions and source galaxy shear in redshift bins $i$ and $j$, $\gamma_t^{ij}(\theta)$, describes the over-density of mass around galaxy positions. The matter associated with the lens galaxy alters the path of the light emitted by the source galaxy, thereby distorting its shape and enabling a non-zero cross-correlation. We consider all possible bin combinations, i.e. allowing the lenses to be in front or behind the sources (in the later case, a non-zero physical signal would be due to magnification). This correlation is also measured in 20 logarithmic angular bins between 2.5 and 250 arcmin. After imposing scale cuts, the total data-vector size in $\gamma_t$ is 304 elements when \maglim is the lens sample and 248 for \redmagic. The validation and robustness of the galaxy galaxy-lensing signal is discussed in detail in \citet*{y3-gglensing}. 

In the Appendix~\ref{sec: DV residuals}, we show the measurements of these two-point functions and compare them with the best-fit \LCDM theory prediction from this work.

\section{Photometric Redshift Calibration}
\label{sec:photoz_calib}

\begin{figure*}
	\begin{center}
		\includegraphics[width=\linewidth]{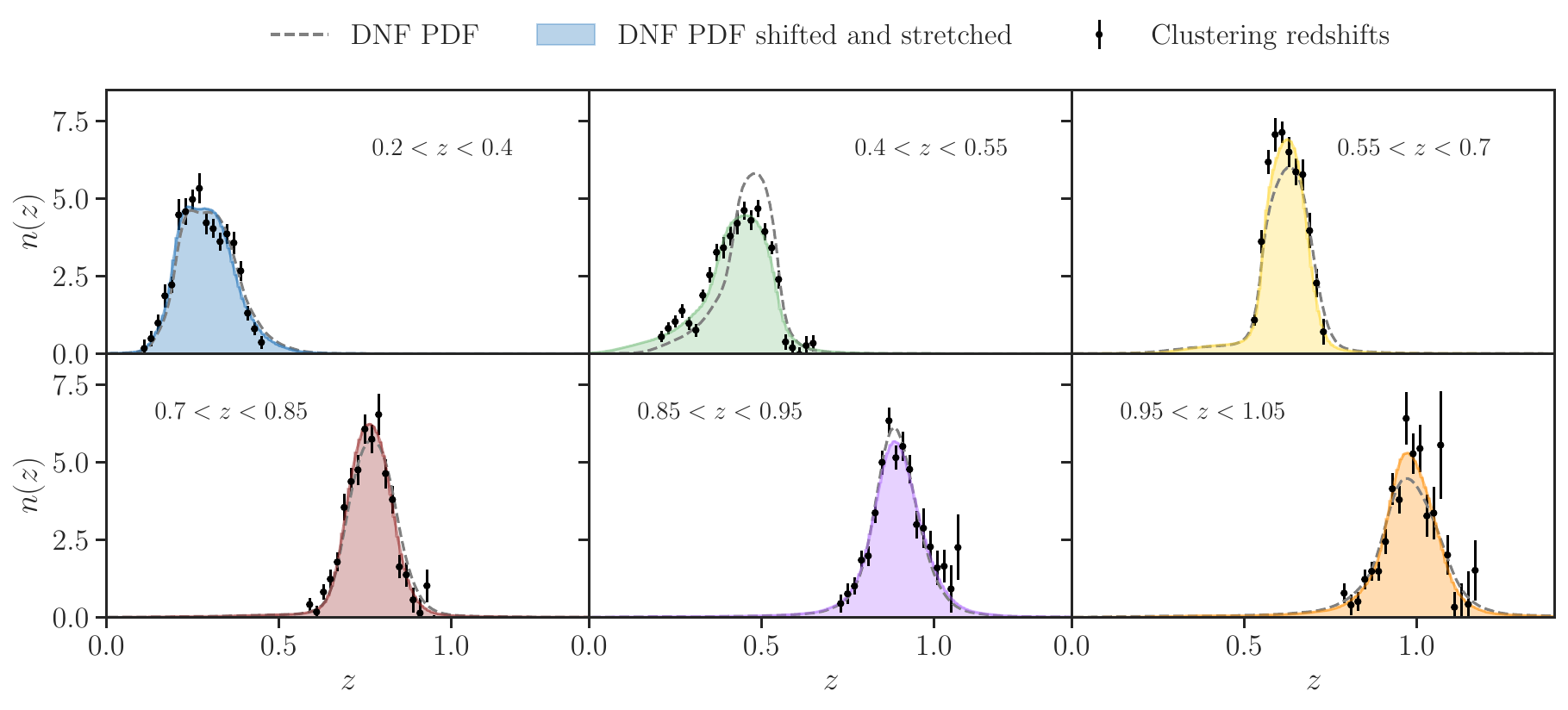}
		\caption{Comparison of $\maglim$ redshift distributions obtained with $\dnf$ (dashed) and clustering redshifts (error bars). The filled regions show the $\dnf$ redshift distributions after applying the fiducial shift and stretch parameters to match the mean and width of the clustering redshift estimates. See Sec.~\ref{subsec:params-and-priors} and \ref{subsec: photo-z parametrization} for the definition and validation of these parameters, respectively. }
		\label{fig: dnf clust-z comp}
	\end{center}
\end{figure*} 

We now present our three different estimations for the true redshift distributions in each tomographic bin and how we cross-validate or combine them.

\subsection{DNF}
\label{sec: dnf}

We use \dnf ~\cite{DNF2016} to select the $\maglim$ galaxies, assign
them into tomographic bins and estimate their redshift distributions
$n(z)$, which are shown in Fig.~\ref{fig: lens source nzs}. For the
former, the algorithm computes a point estimate $z_{DNF}$ of the
true redshift by performing a fit to a hyperplane using 80 nearest neighbors in color and
magnitude space taken from reference set that has an
associated true redshift from a large spectroscopic database. In this work, this database has been constructed using a
variety of catalogs using the DES Science
Portal~\cite{2018A&C....25...58G}.
 The reference catalog includes $\sim 2.2\times10^5$ spectra
matched to DES objects from 24 different spectroscopic catalogs, most
notably SDSS DR14~\cite{sdssdr14}, DES own follow-up through the
OzDES program~\cite{ozdes}, and VIPERS~\cite{VIPERS:2014}. Half of these
spectra have been used as a reference catalog for \dnf.  In addition,
we have added the most recent redshift estimates from the PAU spectro-photometric
catalog (40 narrow bands) from the overlapping CFHTLS W1\footnote{\url{https://www.cfht.hawaii.edu/Science/CFHLS/cfhtlsdeepwidefields.html}} field~\cite{bcnz}. In Appendix~\ref{sec: Comparison of DNF with VIPERS} we compare the \dnf\, point estimates with the spectroscopic redshifts from VIPERS and give more details on the photometric redshift uncertainty of the sample and its outlier rates.

\dnf\, also provides a PDF estimation for each individual galaxy by aggregating the quantities $ z_i = z_{DNF} + s_i$,
where $s_i$ are the residuals resulting from the $i_{th}$ neighbor to
the fitted hyperplane. The sample of all $z_i$ then undergoes a kernel
density estimation process to smooth the distribution.

We then estimate the redshift distribution in each tomographic bin by stacking all the PDFs provided by $\dnf$. These distributions will be calibrated using the cross-correlation
technique (clustering redshifts) described below. Fig.~\ref{fig: dnf clust-z comp} shows that they agree very well with clustering redshifts after such calibration, which consists of applying shift and stretch parameters (see Sec.~\ref{subsec:params-and-priors}) to match the mean and width of the clustering redshift estimates. See Sec.~\ref{subsec: photo-z parametrization} for a detailed description and validation of these parameters.

\subsection{Clustering redshifts}

We calibrate the photometric redshift distributions using clustering redshifts (also known as cross-correlation redshifts) as described in \citet{y3-lenswz}. 
In that work, the angular positions of the \redmagic and \maglim\ galaxies are cross-correlated with a spectroscopic sample of galaxies from the Baryon Oscillation Spectroscopic Survey (BOSS) \cite{boss} and its extension, eBOSS \cite{eboss}. The amplitudes of these cross-correlations are proportional to the redshift overlaps of the photometric and spectroscopic samples. When the spectroscopic sample is divided into small bins, the cross-correlations with each bin put constraints on the true redshift distribution of the photometric samples. Since DES only has partial sky overlap with BOSS and eBOSS, the cross-correlations can only be measured on about $632 \ \text{deg}^2$, or $15 \%$ of the full area.

For this work, the spectroscopic samples are divided into bins of size $dz=0.02$. \citet{y3-lenswz} estimates the DES $n(z)$ in each of these $dz=0.02$ size bins using clustering redshifts across the 5 \redmagic and 6 \maglim\ tomographic bins. 

\subsection{SOMPZ}

\begin{figure*}
	\begin{center}
		\includegraphics[width=\linewidth]{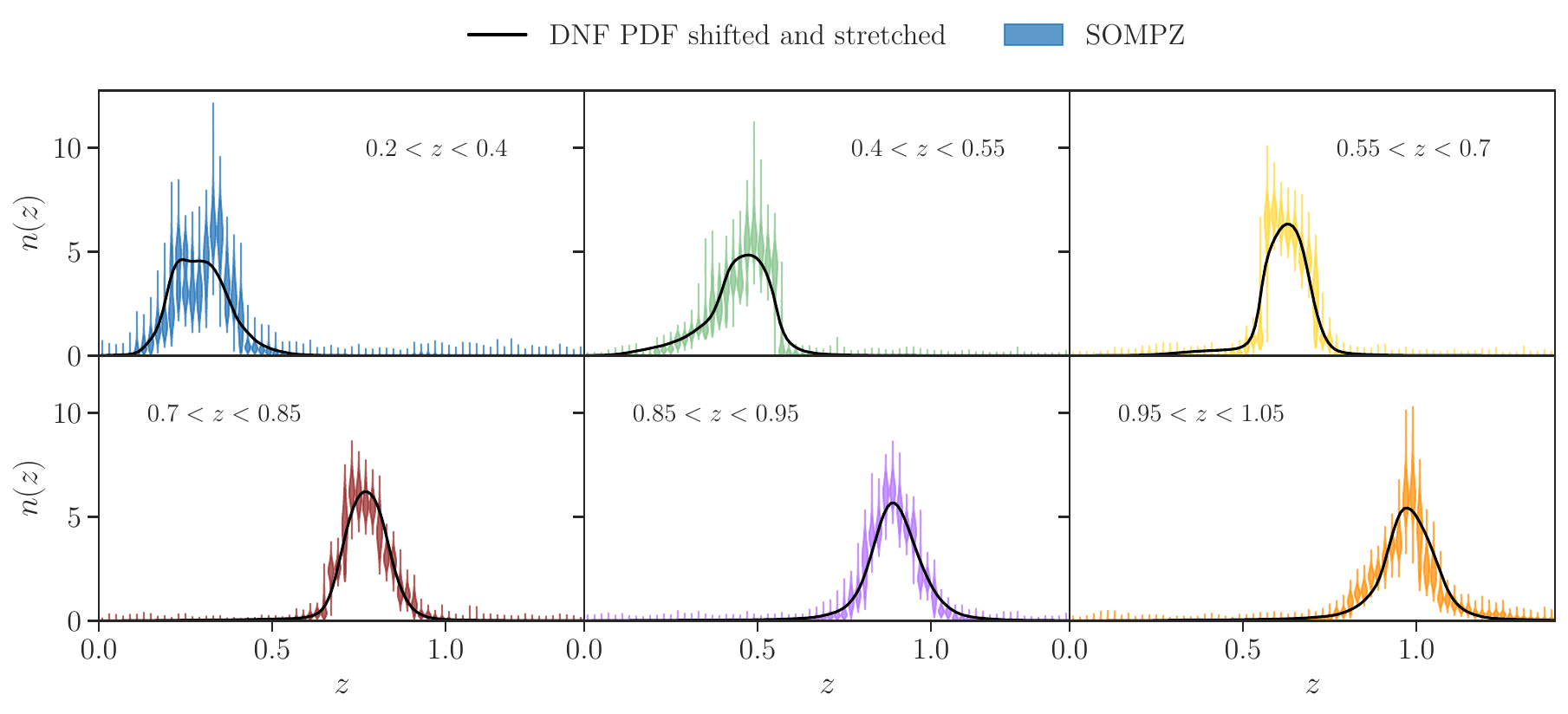}
		\caption{Comparison of $\maglim$ redshift distributions obtained with $\dnf$ (solid black) and SOMPZ (violin plot). The $\dnf$ redshift distributions are shown after applying the fiducial shift and stretch parameters from Table~\ref{tab:params}. }
		\label{fig: dnf sompz comp}
	\end{center}
\end{figure*}

An independent redshift calibration is also performed, analogously to the fiducial method for the source sample \citep{y3-sompz}, placing constraints on the $n(z)$ distribution by relying on the complementary combination of phenotypic galaxy classification done through Self-Organizing Maps (SOMPZ) and the aforementioned clustering redshifts. The methodology and results are described in more detail in \citet{y3-2x2ptaltlenssompz}.
In the SOMPZ method we exploit the additional bands in the DES deep fields to accurately characterize those galaxies, and validate their redshift through three different high precision redshift samples, each of them a different combination of spectra  \citep{2018A&C....25...58G}, PAU+COSMOS \citep{2021MNRAS.501.6103A}, and COSMOS30 \citep{Laigle_2016}.The redshift information is transferred to \maglim through an overlap sample, built by the Balrog algorithm from \citet*{y3-balrog}.

The output of this pipeline is a set of $n(z)$ realizations, whose variability spans all uncertainties. We combine these with clustering redshifts information, estimated in the full redshift range of the BOSS/eBOSS \citep{boss} \citep{eboss} used as reference sample with high quality redshifts, to place a likelihood of obtaining the cross-correlations data given each of the $n(z)$ SOMPZ estimates. The combination places tighter constraints on the shape of the distribution, despite not improving in terms of the uncertainty on the mean of the $n(z)$. 
The final sets of realizations have been computed in bins with $dz=0.02$ and up to $z=3$, and are compatible with the fiducial $\dnf$ n(z), as shown in Fig.~\ref{fig: dnf sompz comp}.

\section{Simulations}
\label{sec:sim}

Parts of the analysis presented in this work have been validated using the \buzzard suite of cosmological simulations. We briefly describe these simulations here and refer the reader to \citet{y3-simvalidation} for a  comprehensive discussion. 

The \buzzard\ simulations are synthetic DES Y3 galaxy catalogs that are constructed from $N$-body lightcones, updated from the version used in the DES Y1 analyses \citep{DeRose2019}. Galaxies are included in the dark-matter-only lightcones using the \textsc{Addgals} algorithm \citep{Wechsler2021, DeRose2021}, which assigns a position, velocity, spectral energy distribution, half-light radius and ellipticity to each galaxy. There are a total of 18  DES Y3 \buzzard\ simulations. Each pair of two Y3 simulations is produced 
from a suite of 3 independent $N$-body lightcones with box sizes of $[1.05,\, 2.6,\, 4.0]\, (h^{-3}\, \rm Gpc^3)$, mass resolutions of $[0.33,\, 1.6, \, 5.9] \, \times10^{11}\, h^{-1}M_{\odot}$, spanning redshift ranges in the intervals $[0.0,\, 0.32,\, 0.84, \,  2.35]$ respectively. These lightcones are produced using the \textsc{L-Gadget2} code, a version of \textsc{Gadget2} \citep{Springel2005} that is optimized for dark--matter-only simulations.
Initial conditions are generated at $z=50$ using \textsc{2LPTIC} \citep{Crocce2012}. Ray-tracing is performed on these simulations using \textsc{Calclens} \citep{Becker2013}, 
with an effective angular resolution of $0.43\, \textrm{arcmin}$. \textsc{Calclens} computes the lensing distortion tensor at each galaxy position
and this is used to calculate angular deflections and rotations, weak lensing shear, and convergence.

The DES Y3 footprint mask is used to apply a realistic survey geometry to each simulation \citep{y3-gold}, resulting in a footprint with an area of 4143.17 $\sqdeg$,
and photometric errors are applied to each galaxy's photometry using a relation derived from \textsc{Balrog} \citep{y3-balrog}.
Weak lensing source galaxies are selected using the PSF-convolved sizes and $i$-band signal--to--noise ratios, matching the non-tomographic source number density in the \metacal source catalog derived from the DES Y3 data. The \textsc{SOMPZ} framework is used to bin source galaxies into tomographic bins, each having a number density of $n_{\rm eff}=1.48$ gal/arcmin$^{2}$,
and to obtain estimates of the redshift distribution of source galaxies \citep{y3-simvalidation,y3-sompz}. The shape noise of the simulations is then matched to that measured in the \metacal catalog per bin.

\begin{figure}
	\begin{center}
		\includegraphics[width=\linewidth]{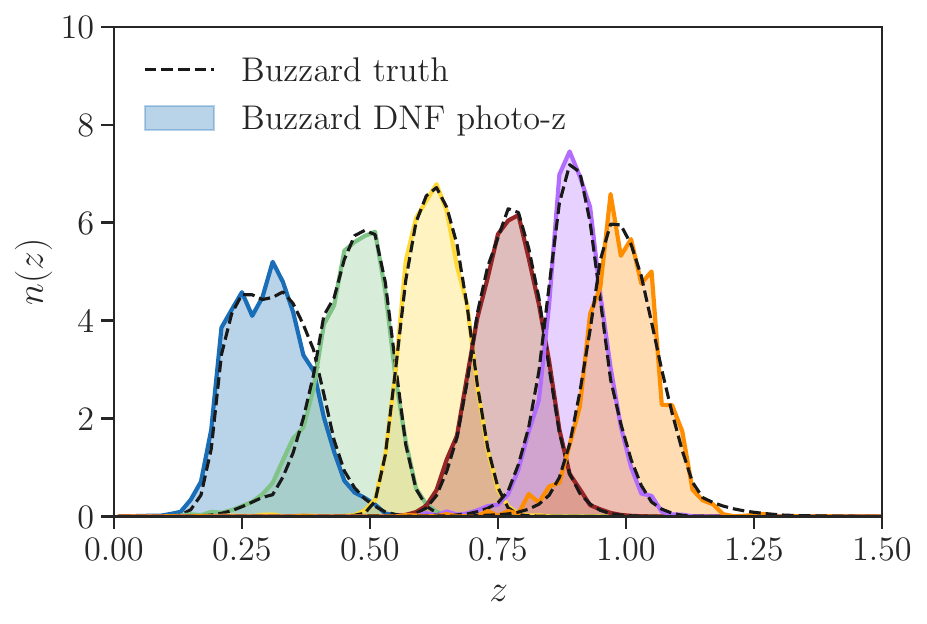}
		\caption{ $\maglim$ Buzzard redshift distributions obtained with $\dnf$ (solid filled) compared with the true distributions (dashed black).}
		\label{fig: buzzard nz comp}
	\end{center}
\end{figure}

In order to reproduce the \maglim sample itself in the simulation, the $\dnf$ code has been run on a subset of one Buzzard realization\footnote{Since running the $\dnf$ code on such a large N-body catalog is computationally expensive, we use only one Buzzard realization to reduce the total running time. }, conservatively cut at \emph{i-band} magnitude $i < 23$ to reduce the running time. Due to the small differences in magnitude/color space between the Buzzard simulation and the DES data, the fiducial \maglim selection applied in Buzzard leads to different number densities and color distributions. When applying the fiducial \maglim selection to Buzzard we find the following number densities $n_{\mathrm{gal}}=[0.106, 0.108, 0.030, 0.072, 0.137, 0.068]$, which should be compared with the values from the data in Table~\ref{tab:samples}. Besides the second bin, which has $n_{\mathrm{gal}}$ very similar to the data, the number densities in Buzzard are about $30\%$ larger in the fifth redshift bin, and between $\sim30\%$ and $70\%$ smaller in the remaining bins.  We therefore re-define an adequate \maglim selection for Buzzard, by identifying the parameters of the linear relation that in each bin minimizes the difference in number density with respect to data, simultaneously requiring the edge values of adjacent bins to correspond, to avoid discontinuity between bins. The resulting linear relation in each bin is quite similar to the data, the larger changes being an increase of around $5\%$ on the $i$-band magnitude cut at $z=0.2$ and $z=0.95$, with smaller differences in the remaining redshift bin edges.   We then estimate the redshift distributions stacking the $\dnf$ nearest-neighbor redshifts (see \ref{sec: dnf}), which is consistent with the fiducial method used for the data. In Fig.~\ref{fig: buzzard nz comp} we compare these with the true redshift distributions, finding good agreement.

The $2\times2$pt data vector is measured without shape noise using the same pipeline as used for the data, with \metacal responses and inverse variance weights set to 1 for all galaxies. 
In Sec.~\ref{sec:scale_cuts} we validate the scale cuts by analyzing these 2$\times$2pt data vectors using the same Buzzard realization for both  \maglim and \redmagic.

\section{Analysis Methodology}
\label{sec:method}

\subsection{Theory modeling}
\label{sec:model}
\subsubsection{Field Level}
\paragraph*{Galaxy Density Field:} On large scales the observed galaxy density contrast is characterised by four main physical contributions, (1) clustering of matter; (2) galaxy bias; (3) redshift space distortions (RSD); and (4) magnification ($\mu$), in such a way that the observed over-density in a tomographic bin $i$ projected on the sky can be expressed as, 
\begin{equation}
\delta_{g,\mathrm{obs}}^{i}(\hat{\mathbf n}) = \delta_{g,\mathrm{D}}^{i}(\hat{\mathbf n}) + \delta_{g,\mathrm{RSD}}^{i}(\hat{\mathbf n}) + \delta_{g,\mu}^{i}(\hat{\mathbf n})\,,
\end{equation}
where the first term is the line-of-sight projection of the three-dimensional galaxy density contrast, 
\begin{equation}
\delta_{g,\mathrm{D}}^{i}(\hat{\mathbf n}) =\int d\chi\, W^i_{\delta}\left(\chi \right)\, \delta_g^{i,(3\mathrm D)}\left(\hat{\mathbf n} \chi, \chi\right)\,,
\end{equation}
with $\chi$ the comoving distance, $W_{\delta}^i = n_g^i(z)\, d z/d\chi$ the normalized selection function of galaxies in tomography bin $i$. 

For the baseline analysis, we adopt a \emph{linear galaxy bias} model with constant galaxy bias per tomographic bin,
\begin{equation}
\delta_g^{i,(3\mathrm D)}(\mathbf x) = b^i \delta_m(\mathbf x)\,.
\end{equation}
Throughout this work, we ignore galaxy bias evolution within a given redshift bin. This assumption is validated with N-body simulations in Sec.~\ref{sec:scale_cuts}.

For some cases, we employ a \emph{perturbative galaxy bias model to third order} in the density field from \cite{Saito_bnl} that includes contributions from local quadratic bias, $b^i_2$, tidal quadratic bias, $b_{s^2}^i$, and third-order nonlocal bias, $b_{3\mathrm{nl}}^i$. As validated in \cite{Pandey2020} and Sec.~\ref{sec:valid}, we fix the bias parameters $b_{s^2}^i$ and $ b_{\rm 3nl}^i $ to their co-evolution value of $b_{s^2}^i=-4(b^i - 1)/7$ and $b_{\rm 3nl}^i=b^i - 1$ \citep{Saito_bnl}.

The magnification term is given by
\begin{equation}
  \delta_{g,\mu}^{i}(\hat{\mathbf n}) = C^i \kappa_{g}^i(\hat{\mathbf n})\,
\label{eq:delta_mu}
\end{equation}
with the magnification bias amplitude $C^i$, and where we have introduce the tomographic convergence field
 \begin{equation}
  \kappa_g^i(\hat{\mathbf n})=\int d\chi \,W_{\kappa,g}^i(\chi)\delta_\mathrm{m}\left(\hat{\mathbf n} \chi, \chi\right)
\end{equation}
 with the tomographic lens efficiency
\begin{equation}
  W_{\kappa,g}^i (\chi)= \frac{3\Omega_{\rm m} H_0^2}{2}\int_\chi^{\chi_H} d\chi' n_g^i(\chi')
 \frac{\chi}{a(\chi)}\frac{\chi'-\chi}{\chi'}\,,
\end{equation}
where $\chi_H$ is the comoving distance to the horizon and $a(\chi)$ is the scale factor. See \citet{y3-generalmethods} for the complete expressions.

\paragraph*{Galaxy Shear Field:} In a similar manner the galaxy shear $\gamma$ has two components and its modeling on large-scales is mainly driven by the following contributions: (1) Gravitational shear, with contributions from dark-matter non-linear growth as well as baryon physics; (2) Intrinsic Alignments (IA); and (3) Stochastic shape noise. 

The two components $\gamma_\alpha$ of the observed galaxy shapes are modeled as gravitational shear ($\mathrm G$) and intrinsic ellipticity. The latter is split into a spatially coherent contribution from intrinsic galaxy alignments (IA), and stochastic shape noise $\epsilon_0$
 \beq
 \gamma_{\alpha}^{j}(\hat{\mathbf n}) = \gamma_{\alpha,\mathrm G}^{j}(\hat{\mathbf n})+\epsilon_{\alpha,\mathrm{IA}}^{j}(\hat{\mathbf n})+\epsilon_{\alpha,0}^j(\hat{\mathbf n})\,.
 \eeq
We model the intrinsic alignments of galaxies using the Tidal Alignment Tidal Torquing model \cite[TATT, ][]{TATT}. This model includes linear aligments with amplitude parameter $a_1$ and redshift evolution parameter $\eta_1$, quadratic alignments with amplitude parameter $a_2$ and redshift evolution parameter $\eta_2$, as well density weighting of the linear alignments with normalization $b_\mathrm{TA}$.
A detailed description of these terms can be found in \cite{y3-generalmethods}, and we refer to \citet*{y3-cosmicshear2} for a discussion of the intrinsic alignment model. For computational convenience, the shear and intrinsic alignment fields are split into E/B-mode components; to leading order in the lensing distortion, B-modes are only generated by intrinsic alignments.

\subsubsection{Two-point statistics}
The observable angular power spectra are then computed by considering the different physical components at the field level. For galaxy-galaxy lensing this results in
\begin{equation}
 C^{ij}_{\delta_{g,\mathrm{obs}}\mathrm{E}}(\ell) = C^{ij}_{\delta_{g,\mathrm{D}}\kappa_{\rm s}}(\ell) + C^{ij}_{\delta_{g,\mathrm{D}}\mathrm{I_E}}(\ell) +  C^{ij}_{\delta_{g,\mu}\kappa}(\ell) + C^{ij}_{\delta_{g,\mu}\mathrm{I_E}}(\ell)\,,
\end{equation}
where we omitted the RSD term, which is negligible for the DES-Y3 lens tomography bin choices \citep{Fang_nonlimber}.
Here, the individual terms are evaluated using the Limber approximation
\begin{align}
     C_{AB}^{ij}(\ell) = \int d\chi \frac{W_A^i(\chi)W_B^j(\chi)}{\chi^2}P_{AB}\left(k = \frac{\ell+0.5}{\chi},z(\chi)\right)\,,
 \end{align}
with $P_{AB}$ the corresponding three-dimensional power spectra, which are detailed in \citet{y3-generalmethods}. 

The angular clustering power spectra has to be evaluated exactly, as the Limber approximation is insufficient at the accuracy requirements of the DES-Y3 analysis. For example, the exact expression for the density-density contribution to the angular clustering power spectrum is
 \begin{align}
  \nonumber   C_{\delta_{g,\rm D}\delta_{g,\rm D}}^{ij} (\ell)=&\frac{2}{\pi}\int d \chi_1\,W^i_{\delta,{g}}(\chi_1)\int d\chi_2\,W^j_{\delta,g}(\chi_2)\\
     &\int\frac{dk}{k}k^3 P_{gg}(k,\chi_1,\chi_2)j_\ell(k\chi_1)j_\ell(k\chi_2)\,,
 \label{eq:Cl-DD}
 \end{align}
and the full expressions including magnification and redshift-space distortion are given in \cite{Fang_nonlimber}. Schematically, the integrand in Eq.~\eqref{eq:Cl-DD} is split into the contribution from non-linear evolution, for which un-equal time contributions are negligible so that the Limber approximation is sufficient, and the linear-evolution power spectrum, for which time evolution factorizes. We use the generalized FFTLog algorithm\footnote{\url{https://github.com/xfangcosmo/FFTLog-and-beyond}} developed in \cite{Fang_nonlimber} to evaluate the full angular clustering power spectrum, including magnification and redshift-space distortion contributions.

The angular correlation functions are then obtained via
\begin{align}
    w^i(\theta) =& \sum_\ell \frac{2\ell+1}{4\pi}P_\ell(\cos\theta) C^{ii}_{\delta_{\mathrm{l,obs}}\delta_{\mathrm{l,obs}}}(\ell)~,\\
    \gamma_t^{ij}(\theta) =& \sum_\ell \frac{2\ell+1}{4\pi\ell(\ell+1)}P^2_\ell(\cos\theta) C^{ij}_{\delta_{\mathrm{l,obs}}\mathrm{E}}(\ell)~,
\end{align}
where $P_\ell$ and $P_\ell^2$ are the Legendre polynomials. 

The tangential shear two-point statistic $\gamma_t$ is a non-local measure of the galaxy-mass cross-correlation, hence the highly non-linear small-scale galaxy mass profile contribute to $\gamma_t$ even at large angular scales. Several methods have been proposed to mitigate this effect \citep[e.g.,][]{Upsilon,PointMass, YStatistic}. Here we adopt analytic marginalization over the mass enclosed below the angular scales included in the analysis. Following the procedure detailed in \citet{PointMass}, we implement the analytic marginalization by modifying the inverse of our fiducial covariance matrix $\mathbf{C}^{-1}$: 
\begin{equation}
\mathbf{C}^{-1}_{\mathrm{wPM}}=\mathbf{C}^{-1} - \mathbf{C}^{-1} \mathbf{U}(\mathbf{I} + \mathbf{U}^{\mathrm{T}}\mathbf{C}^{-1}\mathbf{U})^{-1} \mathbf{U}^{\mathrm{T}}\mathbf{C}^{-1}.
\end{equation}
Here $\mathbf{I}$ is the identity matrix and $\mathbf{U}$ is an $N_{\mathrm{d}} \times n_{\mathrm{lens}}$ matrix, where $N_{\mathrm{d}}$ is the total number of elements in the datavector and $n_{\mathrm{lens}}$ is the number of lens redshift bins. The $i$th column of matrix $\mathbf{U}$ is given by  $\sigma_{B^i}\mathbf{t}^i$, where $\sigma_{B^i}$ is the width of the Gaussian prior on the point-mass parameter $B^i$ we want to marginalize over and $\mathbf{t}^i= \beta^{i,j}/\theta^2$. The $\beta^{i,j}$ parameters modulate the impact of the point-mass parameters on each lens-source pair through their dependence on the effective inverse $\Sigma_\mathrm{crit}$, a geometrical factor that depends on the angular diameter distances to the lenses, to the sources, and between lenses and sources. We refer the reader to \citet{y3-2x2ptbiasmodelling} for a more detailed description of the implementation and validation of the point-mass analytic marginalization.

\subsection{Parameter inference and likelihood}

Parameter inference requires four components: a dataset $\hat{\mathbf D} \equiv\{\hat{w}^i(\theta),\hat{\gamma}_\mathrm{t}^{ij}(\theta)\}$, a theoretical model $\mathbf T_M(\mathbf p) \equiv \{w^i(\theta,\mathbf p),\gamma_\mathrm{t}^{ij}(\theta,\mathbf p)\}$, a description of the covariance of the dataset $\mathbf{C}$, and a set of priors on the model $M$. We assume a Gaussian likelihood
\begin{equation}
\hspace{-0.5em} \ln \mathcal{L}(\hat{\mathbf D}|\mathbf p) = -\frac{1}{2}\left(\hat{\mathbf D}-\mathbf T(\mathbf p)\right)^{\mathrm{T}} \mathbf{C}^{-1}\left(\hat{\mathbf D}-\mathbf T(\mathbf p)\right).
\end{equation}
The covariance is modeled analytically as described and validated in \cite{y3-covariances,CosmoCov,cosmolike,2018MNRAS.479.4998T}. We also account for an additional uncertainty in the $w(\theta)$ covariance that is related to the correction of observational systematics, as described in \citet{y3-galaxyclustering}. The covariance is modified to analytically marginalize over two terms, one given by the difference between correction methods and another one related to the bias of the fiducial correction method as measured on simulations.

In addition to the main galaxy clustering and galaxy-galaxy lensing likelihood described above, we also incorporate small-scale shear ratios (SR) at the likelihood level. This methodology is described in detail by \citet*{y3-shearratio}. The main idea is that by taking the ratio of galaxy-galaxy lensing measurements with a common set of lenses, but sources at different redshifts, the power spectra approximately cancel, and one is left with a primarily geometric measurement. Shear ratios were initially proposed as a probe of cosmology (see e.g. \cite{jain03}), but they have proven more powerful as a method for constraining systematics and nuisance parameters of the model, especially those related to redshift calibration and intrinsic alignments. 

In particular, we choose to use SR on small scales that are not used in the galaxy-galaxy lensing measurement of this work ($<6$ Mpc$/h$, see Sec.~\ref{sec:scale_cuts}), where uncertainties are dominated by galaxy shape noise, such that the likelihood can be treated as independent of that from the galaxy-galaxy lensing data (which also removes small-scale information via point-mass marginalization). As before, we assume a Gaussian likelihood, and derive the analytic covariance matrix from CosmoLike \cite{CosmoCov,cosmolike,2018MNRAS.479.4998T}. Due to the relative lack of signal-to-noise ratio in the higher redshift bins, we use only the three lens bins that are lower in redshift, and compute shear ratios for each lens bin $l$ relative to the fourth source bin, $\gamma_t^{ls}/\gamma^{l4}_t$, $s\in(1,2,3)$. This results in three data vectors per lens bin, or nine overall. See \citet*{y3-shearratio} for the validation and discussion of the SR constraints.

The posterior probability distribution for the parameters is related to the likelihood through the Bayes' theorem:
\begin{equation}
P(\mathbf p|\hat{\mathbf D}, M)\propto \mathcal{L}(\hat{\mathbf D}|\mathbf p, M) \Pi\left(\boldsymbol{\mathrm{p}}|M\right),
\end{equation}
where $\Pi(\boldsymbol{\mathrm{p}}|M)$ is a prior probability distribution on the parameters given a model $M$. We report parameter constraints using the mean of the marginalized posterior distribution of each parameter, along with the 68\% confidence limits (C.L.) around the mean. For some cases, we also report the best-fit maximum posterior values. In addition, in order to compare the constraining power of different analysis scenarios, we use the 2D figure of merit (FoM), defined as $\mathrm{FoM}_{p_1, p_2}=(\det \mathrm{Cov}(p_1, p_2))^{-1/2}$, where $p_1$ and $p_2$ are any two given parameters \cite{Huterer2001,Wang2008}. The FoM is proportional to the inverse area of the confidence region in the $p_1-p_2$ space.

To support redundancy in the likelihood inference we implement two versions of the modeling and inference pipelines: CosmoSIS\footnote{\url{https://bitbucket.org/joezuntz/cosmosis}} \citep{zuntz15} and CosmoLike \citep{cosmolike}. They have been tested against one another to ensure necessary accuracy in calculations of the theoretical two-point functions as described in \citet{y3-generalmethods}. They use a combination of publicly available packages \cite{Lewis:1999bs,CLASSII,Takahashi2012,FastPT} and internally developed code. Parameter inference is primarily performed using the PolyChord sampler \cite{Polychord1,Polychord2}, but results have been cross-checked against Emcee \cite{emcee}.

\subsubsection{Parameter space and priors}
\label{subsec:params-and-priors} 

\begin{table}
	\centering	
	\caption{	\label{tab:params}
	The parameters and their priors used in the fiducial  \maglim $\Lambda$CDM and $w$CDM analyses. The parameter $w$ is fixed to $-1$ in $\Lambda$CDM.  Square brackets denote a flat prior, while parentheses denote a Gaussian prior of the form $\mathcal{N}(\mu,\sigma)$.  \\} 
	\vspace{-0.2cm}
	\begin{tabular}{ccc}
		\hline
		Parameter & Fiducial &Prior\Tstrut\\\hline
		\multicolumn{3}{c}{\textbf{Cosmology}} \Tstrut\\
		$\Omega_{\rm m}$ &  0.3 &[0.1, 0.9] \\ 
		$A_\mathrm{s}10^{9}$ & 2.19 & [$0.5$, $5.0$]  \\ 
		$n_{\rm s}$ & 0.97 & [0.87, 1.07]  \\
		$w$ &  -1.0  &[-2, -0.33]   \\
		$\omb$ & 0.048 &[0.03, 0.07]  \\
		$h_0$  & 0.69  &[0.55, 0.91]   \\
		$\Omega_\nu h^210^3$ & 0.83 & [0.6, 6.44]
		\\\hline

		\multicolumn{3}{c}{\textbf{Linear galaxy bias  } }\Tstrut	 \\
		$b_{i}$  & $1.5, 1.8, 1.8, 1.9, 2.3, 2.3$ & [0.8,3.0]\\\hline
		
		\multicolumn{3}{c}{\textbf{ Non-linear galaxy bias  } }\Tstrut	 \\
		$b_1^{i}\sigma_8 $  & $1.43, 1.43, 1.43, 1.69, 1.69, 1.69 $ & [0.67,3.0]\\
		$b_2^{i}\sigma_8^2 $  & $0.16, 0.16, 0.16, 0.36, 0.36, 0.36 $ & [-4.2, 4.2]\\\hline
		
		\multicolumn{3}{c}{\textbf{Lens
				magnification } } \Tstrut \\
		$C_{i} $ & 0.43, 0.30, 1.75, 1.94, 1.56, 2.96 & Fixed\\ \hline

		\multicolumn{3}{c}{\textbf{Lens photo-z shift }}	\Tstrut \\
		$\Delta z^1_{\rm l}$ & -0.009 & ($
		-0.009, 0.007$)\\ $\Delta z^2_{\rm l}$ & -0.035 & ($
		-0.035, 0.011$)\\ $\Delta z^3_{\rm l}$ & -0.005 & ($
		-0.005, 0.006$)\\ $\Delta z^4_{\rm l}$ & -0.007 & ($
		-0.007, 0.006$)\\ $\Delta z^5_{\rm l}$ & 0.002 & ($
		0.002, 0.007$)\\ $\Delta z^6_{\rm l}$ & 0.002 & ($
		0.002, 0.008$)\\ $\sigma z^1_{\rm l}$ & 0.975 & ($
		0.975, 0.062$)\\ $\sigma z^2_{\rm l}$ & 1.306 & ($
		1.306, 0.093$)\\ $\sigma z^3_{\rm l}$ & 0.870 & ($
		0.870, 0.054$)\\ $\sigma z^4_{\rm l}$ & 0.918 &
		($0.918, 0.051$)\\ $\sigma z^5_{\rm l}$ & 1.080 &
		($1.08, 0.067$)\\ $\sigma z^6_{\rm l}$ & 0.845 &
		($0.845, 0.073$) \\\hline 
		
		\multicolumn{3}{c}{{\bf
				Intrinsic alignment}}\Tstrut \\
		$a_{i}$ ($i\in [1,2]$)   & 0.7, -1.36 &  [$-5,5$ ]\\
		$\eta_{i}$  ($i\in [1,2]$) & -1.7, -2.5  & [$-5,5$ ]\\
		$b_{\mathrm{TA}}$   & 1.0  & [$0,2$] \\
		$z_0$ & 0.62   &  Fixed\\
		\hline
		\multicolumn{3}{c}{{\bf Source \photoz}}\Tstrut \\
		$\Delta z^1_{\rm s}$  & 0.0  & ($0.0, 0.018$) \\
		$\Delta z^2_{\rm s}$  & 0.0  & ($0.0, 0.013$) \\
		$\Delta z^3_{\rm s}$  & 0.0  & ($0.0, 0.006$) \\
		$\Delta z^4_{\rm s}$  & 0.0  & ($0.0, 0.013$) \\
		\hline
		\multicolumn{3}{c}{{\bf Shear calibration}}\Tstrut \\
		$m^1$ & 0.0  & ($-0.006, 0.008$)\\
		$m^2$ & 0.0  & ($-0.010, 0.013$)\\
		$m^3$ & 0.0  & ($-0.026, 0.009$)\\
		$m^4$ & 0.0  & ($-0.032, 0.012$)\\
		\hline
	\end{tabular}
\end{table}

We sample the likelihood of clustering and galaxy-galaxy lensing measurements over a set of cosmological, astrophysical and systematics parameters, whose fiducial values and  priors are summarized in Table \ref{tab:params}. We do this in two cosmological models, \LCDM and \wCDM, in both cases assuming a flat universe and a free neutrino mass.

{\it Cosmological parameters:} For \LCDM we sample over the total matter density $\Omega_m$, the amplitude of primordial scalar density fluctuations $A_s$, the spectral index $n_s$ of their power spectrum, the
baryonic density  $\Omega_b$ and the Hubble parameter $h$. We also vary the massive neutrino density $\Omega_{\nu}$ through the combination $\Omega_{\nu} h^2$. In \wCDM this list is extended to include a free parameter $w$ for the equation of state of dark energy. In both models, flatness is imposed by setting $\Omega_\Lambda = 1 - \Omega_m$.  The prior ranges for these parameters are set such that they encompass five times the $68\%$ C.L. from external experiments in the case they are not strongly constrained by DES itself. 
In all, we consider six parameters in \LCDM and 7 in \wCDM. 
Instead of quoting constraints on $A_s$, we will refer to the rms amplitude of mass fluctuations on 8$\mpc$ scale in linear theory, $\sigma_8$, or the related parameter $S_8$, 
\begin{equation}
S_8 \equiv \sigma_8 \left( \dfrac{\Omega_m}{0.3} \right)^{0.5},
\end{equation}
which typically is better constrained because it is less correlated with $\Omega_m$.

{\it Astrophysical parameters:} In addition we marginalize over a number of parameters related to the galaxy biasing model, the intrinsic alignment model, and the magnification model. For linear galaxy bias we include one free parameter per bin $b^i$, or two in the case of non-linear bias $b_1^i - b_2^i$. Tidal galaxy biases are kept fixed to their local Lagrangian expressions in terms of the linear bias, as discussed in Sec.\ref{sec:model} and \citet{Pandey2020}. Our baseline model for intrinsic alignment of galaxies (TATT, \cite{TATT}) is parameterized by an amplitude $a_i$ and a power law index $\eta_i$ for both the tidal alignment and the tidal torque terms, in addition to a global source galaxy bias parameter $b_{\rm TA}$. Lastly we consider one parameter per lens tomographic bin to account for the amplitude of lens magnification. This parameter is however kept fixed in our baseline analysis, to the value calibrated on realistic simulations, as described in \citet*{y3-2x2ptmagnification}. 

{\it Systematic parameters:} Photometric redshift systematics are parameterized by an additive shift to the mean redshift of each bin, given by $\Delta z_{\rm l}$ for lenses and $\Delta z_{\rm s}$ for sources, such that
\begin{equation}
n^i(z) = n^i_{pz}(z-\Delta z^i).
\label{eq: delta-z}
\end{equation}
\citet*{y3-hyperrank} demonstrated that our uncertainties in higher order modes of the source $n(z)$'s, besides the mean redshift, have negligible impact in cosmological constrains from cosmic shear. Hence the above treatment is sufficient. For the lenses, however, we have found that current uncertainties on the shape of the $n(z)$'s are important, in part because the clustering and galaxy-galaxy lensing kernels are more localized. Thus, in the case of the lenses, we also parameterize the uncertainty on the width of the redshift distribution by a stretch $\sigma z$, such that
\begin{equation}
n^i(z) = \dfrac{1}{\sigma z^i}n^i_{pz}\left(\dfrac{z-\langle z\rangle}{\sigma z^i}+\langle z\rangle\right).
\label{eq: stretch pz}
\end{equation}

The priors for the lens shift and stretch parameters were calibrated in  \citet{y3-lenswz} and are specified in Table \ref{tab:params}. In Sec.~\ref{subsec: photo-z parametrization} we validate this parameterization for $\maglim$, showing that it allows us to recover unbiased cosmology and galaxy bias values. See \cite{y3-lenswz} for the equivalent validation of $\redmagic$ shift and stretch parameters and Table I in \cite{y3-2x2ptbiasmodelling} for further details (note that in the case of $\redmagic$ the stretch parameter is only required for the highest tomographic bin).

In addition, the measured ellipticity is a biased estimate of the underlying true shear. This bias is taken into account in our pipeline through a multiplicative bias $m$ correction, as defined in Eq.~\eqref{eqn:shearbias}. This correction is applied as an average over all the galaxies in each source tomographic bin. The priors on these parameters are derived in \cite{y3-imagesims} using image simulations and are listed in Table~\ref{tab:params}. \citet{y3-imagesims} founds the additive bias $c$ to be negligible compared to the multiplicative bias.

In total, our baseline likelihood analysis (i.e. linear galaxy bias) marginalizes over 37 parameters in \LCDM (or 38 in $w$CDM). The extension to non-linear galaxy bias adds one parameter per lens bin.

\subsection{Blinding}

We protect our results against observer bias by systematically shifting our results in a random way at various stages of the analysis \citet*{y3-blinding} to prevent us from knowing the true cosmological results or model fit until all decisions about the analysis have been made. This process and the decision tree to unblind is described in detail in \cite{y3-3x2ptkp}. We describe some changes that were made post-unblinding in Sec.~\ref{sec:unblinding}.

\subsection{Quantifying internal consistency}
\label{subsec:ppds}

We defined a process before unblinding for testing internal consistency of the data based on the Posterior Predictive Distribution (PPD). We can derive a probability to exceed $p$ from this process, which either tells us $p$ of a dataset given a chosen model (like $\Lambda$CDM) and covariance or the $p$ of a dataset given constraints on the model from a different potentially correlated dataset. We have verified the consistency of the final 2$\times$2pt data vector to the \LCDM and $w$CDM models and its consistency with our cosmic shear data in \cite{y3-cosmicshear1,y3-cosmicshear2}. These PPD statistics are validated in  \citet*{y3-inttensions}.

\section{Model Validation}
\label{sec:valid}

In this section, we validate our modeling pipeline using both simulated noiseless theory data vectors and measurements from the Buzzard N-body simulations. We quantify the impact of several systematic effects that are not included in our baseline model (Sec.~\ref{sec:model}) in Sec.~\ref{sec:scale_cuts} and validate the parametrization choice for systematics effects that are included in the baseline model in Sec.~\ref{sec:stress_tests}. In Sec.~\ref{subsec: photo-z parametrization} we validate our parametrization and priors for the uncertainties in the \maglim redshift distributions.

The validation procedure is the same for both Sec.~\ref{sec:scale_cuts} and \ref{sec:stress_tests}: we generate a synthetic data vector including a variation around the baseline model, which is then analyzed with the baseline model. We carry out simulated (cosmic shear, 2x2pt, 3x2pt, 3x2pt+\textit{Planck}) analyses in $\Lambda$CDM and $w$CDM on contaminated data vectors and quantify the 2D parameter bias in $(\Omega_{\mathrm m},S_8)$ for $\Lambda$CDM and in $(\Omega_{\mathrm m},w)$ for $w$CDM. As described in \citet{y3-generalmethods}, we require $\Delta \chi^2 < 1$ and the 2D parameter biases to be smaller than $0.3\sigma_{2\mathrm{D}}$ for each variation study, in order to ensure that the total potential systematic bias is well below $1\sigma$ statistical uncertainty.  

\subsection{Scale cuts}
\label{sec:scale_cuts}
The baseline model summarized in Sec.~\ref{sec:model} is incomplete in modeling astrophysical effects, with the leading unaccounted systematics being non-linear galaxy bias and the impact of baryonic physics on the matter power spectrum. We thus design scale cuts that remove the data points most affected by non-linear galaxy bias and baryonic feedback and control systematic biases in the inferred cosmological parameters to less than $0.3\sigma_{2\mathrm{D}}$.

Of these two effects, baryonic feedback is the dominant for small-scale cosmic shear modeling (see e.g. \cite{Semboloni2011,Chisari2019}), while non-linear bias  is the dominant contamination for galaxy clustering and galaxy-galaxy lensing. Hence we review here only the procedure for mitigating the effect of non-linear bias through scale cuts, and refer to \cite{y3-generalmethods, y3-cosmicshear1,y3-cosmicshear2} for the discussion of scale cuts mitigating baryonic effects on cosmic shear. We note that the scale cut analysis also includes baryonic effects on the matter power spectrum in galaxy clustering and galaxy-galaxy lensing, however these contaminants are far subdominant compared to non-linear bias effects and we refer to \cite{y3-generalmethods} for details.

We employ the perturbative galaxy bias model summarized in Sec.~\ref{sec:model} to compute synthetic data vectors that include contributions from non-linear galaxy bias to galaxy clustering and galaxy-galaxy lensing. The fiducial parameter values for $b_2^i$ used to compute the contaminated data vector are based on bias measurements for a \maglim-like sample selection in mock catalogs \cite{Pandey2020}.
We then run simulated cosmology analyses for a set of scale cut proposals that vary the minimum comoving transverse scale $R_{\rm{min}}$ included in the analysis, corresponding to an angular scale cut
$
\theta_{\mathrm{min}}^i  =R_{\mathrm{min},w/\gamma_{\rm{t}}}/\chi(z^i)
$
for each lens tomography bin $i$. We find that
\begin{equation}
\left(R_{\mathrm{min},w},\,R_{\mathrm{min},\gamma_{\rm{t}}}\right) =\big(8\, \mathrm{Mpc}/h,\,6\, \mathrm{Mpc}/h \big)\,
\end{equation}
meets out requirements. As mentioned in Sec.~\ref{sec:model}, $\gamma_t$ is a non-local quantity, and therefore its value at any angular scale $\theta$ carries information from all the scales smaller than $\theta$. For this reason, the scale cuts in the DES Y1 3$\times$2pt analysis were larger for $\gamma_t(\theta)$ ($12\,  \mathrm{Mpc}/h$) than for $w(\theta)$ ($8\, \mathrm{Mpc}/h$). Here, we are able to include smaller scales in our analysis of galaxy-galaxy lensing (up to $6\, \mathrm{Mpc}/h$) thanks to the point-mass analytic marginalization procedure \cite{PointMass} that we adopt to mitigate the impact of this non-locality. See Sec.~\ref{sec:model} and \citet{Pandey2020} for a more detailed description of this method.

\begin{figure}
	\begin{center}
		\includegraphics[width=\linewidth]{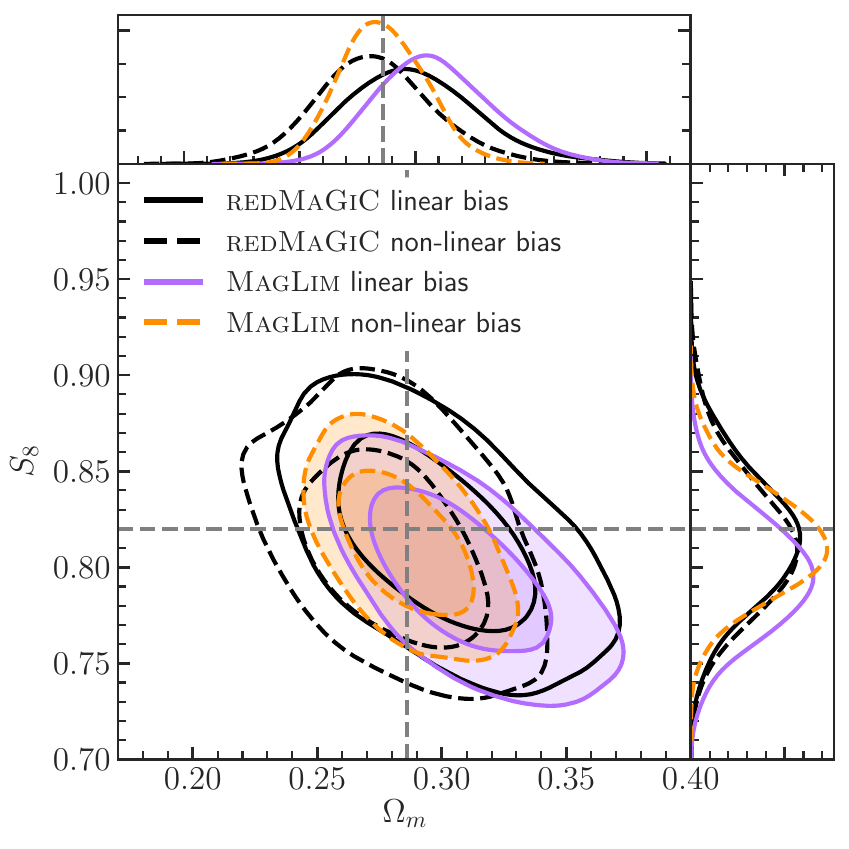}
		\caption{Marginalized 2$\times$2pt constraints on $S_8$ and $\Omega_m$ in \LCDM using measurements on one Buzzard realization (see Sec.~\ref{sec:sim}). The dashed gray lines indicate the true values assumed in the simulation.  These results validate the assumed scale cuts for linear and non-linear galaxy bias modeling, as described in Sec.~\ref{sec:scale_cuts}.  }
		\label{fig:sim_constraints}
	\end{center}
\end{figure}

The same non-linear galaxy bias model is also used in the analysis pipeline, which allows us to include more non-linear scales than the linear bias analyses. We base the scale cut validation for the non-linear bias model on mock catalogs, as extending the scale cut procedure described above to validate the third-order bias model would require higher-order perturbative modeling.  \citet{Pandey2020} compared the non-linear galaxy bias model predictions to 3D matter-galaxy and galaxy-galaxy correlation function measurements from mock catalogs, and found few-percent level accuracy down to $4\, \mathrm{Mpc}/h$. 

For the \redmagic sample, the systematic bias on cosmological parameters is characterized in \citet{y3-simvalidation} using a suite of 18 DES-Y3 mock realizations; their analysis found the linear bias model with $\big(8\, \mathrm{Mpc}/h,\,6\, \mathrm{Mpc}/h \big)$ scale cuts and non-linear bias with $\big(4\, \mathrm{Mpc}/h,\,4\, \mathrm{Mpc}/h \big)$ to be sufficiently accurate for the accuracy of DES-Y3 analyses. See also \citet{y3-2x2ptbiasmodelling} for further details on the non-linear bias validation.  

As there is only one DES-Y3 mock realization available for the \maglim\ sample (described in Sec.~\ref{sec:sim}), we compare parameter inferences from \redmagic and \maglim analyses based on the same realization, which are shown in Fig.~\ref{fig:sim_constraints}. We note that the assumed scale cuts in terms of $\mathrm{Mpc}/h $ are the same for both samples. We find parameter shifts between \redmagic and \maglim baseline analyses of the same realization at the level of $0.16\sigma_{2\rm{D}}$ (linear bias) and $0.05\sigma_{2\rm{D}}$ (non-linear bias), and between baseline (linear bias) and non-linear bias analyses of the \maglim realization at the level of $0.45\sigma_{2\rm{D}}$. The latter is similar to the \redmagic  parameter shift  between linear and non-linear bias analyses in Fig.~\ref{fig:sim_constraints}, which is at the level of  $0.5\sigma_{2\rm{D}}$. These shifts are larger than our threshold of $0.3\sigma_{2\mathrm{D}}$, however this is due to the statistical noise in the Buzzard data vector, since it is measured from just one realization. When using the mean of the 18 mock realizations \cite{y3-simvalidation}, this statistical noise is reduced and we find a shift of $0.01\sigma_{2\rm{D}}$ between \redmagic linear and non-linear bias analyses.  We hence conclude that the linear and non-linear bias scale cuts are sufficient for the analyses presented in this paper. 

Our modeling assumption of constant galaxy bias parameters in each redshift bin is validated as well through the analysis of the $\maglim$ Buzzard mock catalog, which contains non-parametric redshift evolution and non-linear evolution of the matter field. If this assumption were insufficient, we would obtain cosmological constraints biased from the true values in Fig.~\ref{fig:sim_constraints}. Therefore, the results from Fig.~\ref{fig:sim_constraints}, which show that we recover the true cosmology, validate that systematic biases due to galaxy bias redshift evolution are insignificant for the DES-Y3 analysis.

\subsection{Model stress tests}
\label{sec:stress_tests}
\begin{table}
\centering
\caption{Parameter biases from model stress tests, c.f.~Sec.~\ref{sec:stress_tests} for details. The second and third columns report two-dimensional parameter biases relative to the 2D-marginalized parameter uncertainty of the baseline analysis, in $\Lambda$CDM and $w$CDM, respectively. The results for 3$\times$2pt are also shown for completeness. }
\label{tab:stress_test}
\vspace{0.15cm}
\begin{tabular}{l |r | r}
Model stress test & $\Delta_{2\mr{D}}$ ($\Omega_{\mathrm m}, S_8$) & $\Delta_{2\mr{D}}$  ($\Omega_{\mathrm m}, w$)\\
\hline\Tstrut
matter power spectrum, 2$\times$2pt & $<0.01 \sigma_{2\mr{D}}$ &$<0.01 \sigma_{2\mr{D}}$\\
matter power spectrum, 3$\times$2pt & $0.05 \sigma_{2\mr{D}}$& $0.01 \sigma_{2\mr{D}}$\\
\hline\Tstrut
higher-order lensing, 2$\times$2pt &$0.12 \sigma_{2\mr{D}}$ & $<0.01 \sigma_{2\mr{D}}$\\
higher-order lensing, 3$\times$2pt & $0.26\sigma_{2\mr{D}}$ & $<0.01 \sigma_{2\mr{D}}$\\
\hline
\end{tabular}
\end{table}

The baseline analysis model requires several parameterization \emph{choices} to evaluate angular correlation functions. These parameterization choices are practical but imperfect approximations, and we need to demonstrate their robustness for the DES-Y3 analyses. In practice, we stress-test these baseline parameterization choices by generating simulated theory data vectors using alternate parameterizations, which are then analyzed with the baseline model. A comprehensive overview of model choices and their validation is given in \cite{y3-generalmethods}, we highlight here the systematics most relevant to the analysis presented in this paper.

\paragraph*{Matter power spectrum} The baseline model adopts the \citet{Takahashi12} recalibration of the \textsc{halofit} fitting formula \citep{halofit} for the gravity-only matter power spectrum, including the \citet{Bird_halofit} prescription for the impact of massive neutrinos. In order to quantify the model accuracy we compare \textsc{halofit} against more recent matter power spectrum emulators, which are based on larger simulations and thus more accurate. Specifically, we compare to the \textsc{Euclid Emulator} \citep{emu_euclid}. The results of this test are summarized in Table~\ref{tab:stress_test}, we find the systematic parameter biases due to non-linear power spectrum modeling to be insignificant for the DES-Y3 analysis. The results for 3$\times$2pt are also shown for completeness.

\paragraph*{Magnification} The magnification coefficients $C^{i}$ in Eq.~\eqref{eq:delta_mu} are fixed in this analysis to values derived in \cite{y3-2x2ptmagnification}. \citet*{y3-2x2ptmagnification} demonstrates that DES-Y3 cosmology constraints are robust to biases in the estimated values for the magnification coefficients, including the extreme scenario of ignoring lens magnification in the analysis. We leave tests of the redshift evolution of these coefficients to future analyses. 
\paragraph*{Higher-order lensing effects} While the baseline analysis model includes weak lensing to first order in the distortion tensor, \emph{reduced shear} \cite{Dodelson06,Shapiro09} and \emph{source magnification} \cite{Peter_Bmodes,lensingbias} contribute to the angular correlation functions at next-to-leading order. We compute these corrections in \cite{y3-generalmethods}. As described above, we generate a theory data vector with these corrections and analyze it with the baseline analysis model to test the robustness of the constraints. The results of this test for \maglim,  summarized in Table~\ref{tab:stress_test}, show that the systematic parameter biases due to higher-order weak lensing effects are insignificant for the DES-Y3 analysis. 

  \begin{figure*}
	\begin{center}	
		\includegraphics[width=\linewidth]{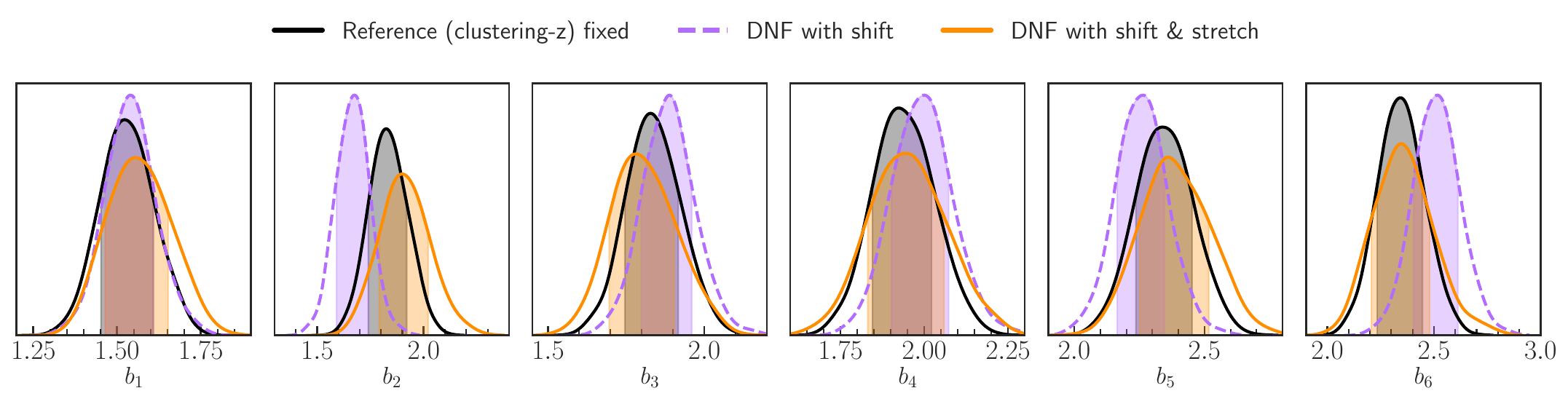}
		\caption{Marginalized 1D constraints on the galaxy bias parameters in $\Lambda$CDM  from  2$\times$2pt simulated theory data vectors using different $n(z)$ estimates. We use clustering redshifts as reference (solid black) and compare the contours with $\dnf$ marginalizing over $\Delta z$ shifts (purple dashed) and both shift and stretch parameters (orange filled). }
		\label{fig: pz validation cosmo}
	\end{center}
\end{figure*}

\begin{figure}
	\begin{center}
		\includegraphics[width=\linewidth]{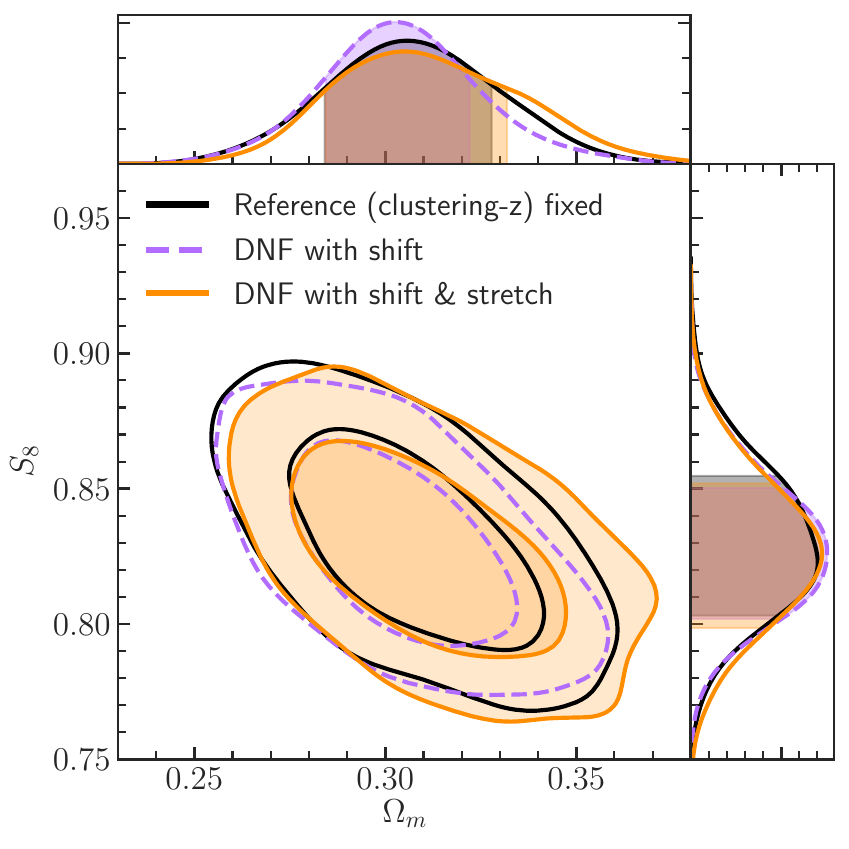}
		\caption{Marginalized 2$\times$2pt constraints on $S_8$ and $\Omega_m$ in $\Lambda$CDM  from simulated theory data vectors using different $n(z)$ estimates. We use clustering redshifts as reference (solid black) and compare the contours with $\dnf$ marginalizing over $\Delta z$ shifts (purple dashed) and both shift and stretch parameters (orange filled). }
		\label{fig: pz validation bias}
	\end{center}	
\end{figure}

\subsection{Photometric redshift parametrization}
\label{subsec: photo-z parametrization}

As described in Sec.~\ref{subsec:params-and-priors}, we parameterize the uncertainty in the mean and width of the redshift distributions by introducing an additive \emph{shift} and a \emph{stretch} parameter for each tomographic bin. In this section, we describe the calibration and validation of these parameters. 

We calibrate the $\dnf$ photometric redshift distributions via a two-parameter $\chi^2$ least squares fit to the clustering redshift estimate of $n(z)$ from \citet{y3-lenswz} (see also Fig.~\ref{fig: dnf clust-z comp}), which we denote as $n_{\rm wz}(z)$. The functional form of this fit is given by the combination of Eqs.~\eqref{eq: delta-z} and \eqref{eq: stretch pz} : 
\begin{equation}
\label{wz fit}
n^i_{\rm wz}(z) = \dfrac{1}{\sigma z^i}n^i_{pz}\left(\dfrac{z-\langle z\rangle-\Delta z^i}{\sigma z^i}+\langle z\rangle\right).
\end{equation}
 where $\langle z\rangle$ is the mean redshift of the initial $\dnf$ photometric redshift distribution, $n_{\rm pz}(z)$.
 
 Additionally, we consider an approach with only shift parameters (i.e. neglecting the uncertainty on the widths of the distributions). Similarly, we calibrate the shift parameters by doing a $\chi^2$ fit to the clustering-redshifts estimate $n_{\rm wz}(z)$ with Eq.~\eqref{eq: delta-z}. The details of this procedure are described in \cite{y3-lenswz}. 
 
 The $\chi^2$ fits provide an estimate on the priors of these parameters. We then validate these priors by ensuring that they allow us to recover unbiased cosmological constraints and galaxy bias values. For this purpose, we use a simulated theory data vector generated with the clustering redshifts distributions $n_{\rm wz}(z)$. Since the tails of $n_{\rm wz}(z)$ are noisy and can have negative values, we cut the tails following \cite{y3-lenswz}. Additionally, as noise in the clustering-redshift point estimates could induce biases in the cosmology, we use a smoothed version of $n_{\rm wz}(z)$  that is the result of fitting a combination of gaussians to the individual points.
 
We use as reference a simulated analysis with $n_{\rm wz}(z)$ and no marginalization of photometric redshift parameters, and then we analyze the simulated theory data vector generated with $n_{\rm wz}(z)$ using the $\dnf$ $n(z)$ marginalizing over shifts or both shift and stretch parameters. The results are shown in Figs.~\ref{fig: pz validation cosmo} and \ref{fig: pz validation bias}. We find that, while marginalizing over the shifts allows us to recover the reference cosmology, it underestimates the contours and fails to recover the galaxy bias parameters in all the bins. When considering the uncertainty on the width through the stretch parameters, we recover both the reference cosmology and galaxy bias values. Hence we conclude that varying both shift and stretch parameters is needed for the $\maglim$ sample, and that the priors estimated in \cite{y3-lenswz} and listed in Table~\ref{tab:params} are sufficient.

\section{Results}
\label{sec:cosmology-constrains}

\subsection{Unblinding}
\label{sec:unblinding}

After passing all the tests outlined in \cite{y3-3x2ptkp} and summarized in Sec.~\ref{subsec:ppds}, we unblinded the \maglim\ results.  We then updated the covariance matrix so that its elements were computed at the best fit parameter values found in the $3\times2$pt unblinded run. The clustering parameters (galaxy bias $b$ multiplied by $\sigma_8$) were found to be smaller in the unblinded result compared to the original parameters assumed in the covariance computation, and therefore the updated covariance matrix assumed less clustering. Since the error bars on $w(\theta)$ on large scales are dominated by cosmic variance, they were considerably smaller ($\sim 50\%$) in the new covariance matrix. We then reran all the chains and discovered that the fit to all cosmological models considered in this work was poor. With the old covariance matrix, our analysis passed our requirement on goodness-of-fit for unblinding ($p>0.01$). However, the best-fit $\Lambda$CDM model with the new covariance matrix had a PPD goodness-of-fit $p< 10^{-3}$.

The problem was localized to be related to the two highest redshift lens bins. We include a comprehensive discussion of our tests after unblinding in the Appendix~\ref{sec: appendix unblinding results}, and summarize here our conclusions. We found that the model struggled to provide a consistent fit to both galaxy clustering and galaxy-galaxy lensing amplitudes on the last two tomographic bins. One way to demonstrate this is to allow the galaxy bias in clustering to differ from the bias in galaxy-galaxy lensing with the ratio in the $i^{th}$ lens bin given by a parameter $X_i$. On large scales where the linear bias assumption is valid, $X_i$ are expected to be equal to unity at the percent level (see e.g. \cite{Pen1998}). This is true for the lowest 4 bins, but in the last two bins we obtain $X_5=0.77\pm 0.10$ and $X_6=0.54\pm0.07$ when combining the 2$\times$2pt data vector with cosmic shear (see equivalent results at fixed cosmology in \citet{y3-2x2ptbiasmodelling}). 
These values of $X$ in the highest redshift bins, the bins in which the redshifts of galaxies are most difficult to determine, pointed to problems with these two bins. 
If we had been running with a more appropriate covariance matrix pre-unblinding, we would very likely have made the decision to drop the last 2 bins. Therefore, in what follows, we present results using only the four lowest redshift bins. This results in an appropriate model fit to both \LCDM and $w$CDM, with $p=0.02$. We have not identified yet a specific systematic origin for these issues, but a calibration problem for high photometric redshifts (where the available spectroscopic references are extremely sparse) is highly plausible.

We present more details in the Appendix~\ref{sec: appendix unblinding results}; see Fig.~\ref{fig: cov comparison lcdm}. Without the highest two redshift bins, the cosmological constraints get weaker as expected, but using the full set of six bins is not justified given the bad fit and the internal inconsistency between clustering and galaxy-galaxy lensing, that clearly points to a systematic effect. There is also a shift in the best fit value of the parameters, but this is at the 1/2-sigma level for both $S_8$ and $\Omega_m$, and clearly within the statistical uncertainty when using the first four redshift bins. Furthermore, the parameter shift due to dropping the highest two redshift bins is significantly reduced when combined with cosmic shear data~\cite{y3-3x2ptkp}.

\subsection{$\Lambda$CDM}
\label{sec:LCDM}
 
\begin{figure}
	\begin{center}
		\includegraphics[width=\linewidth]{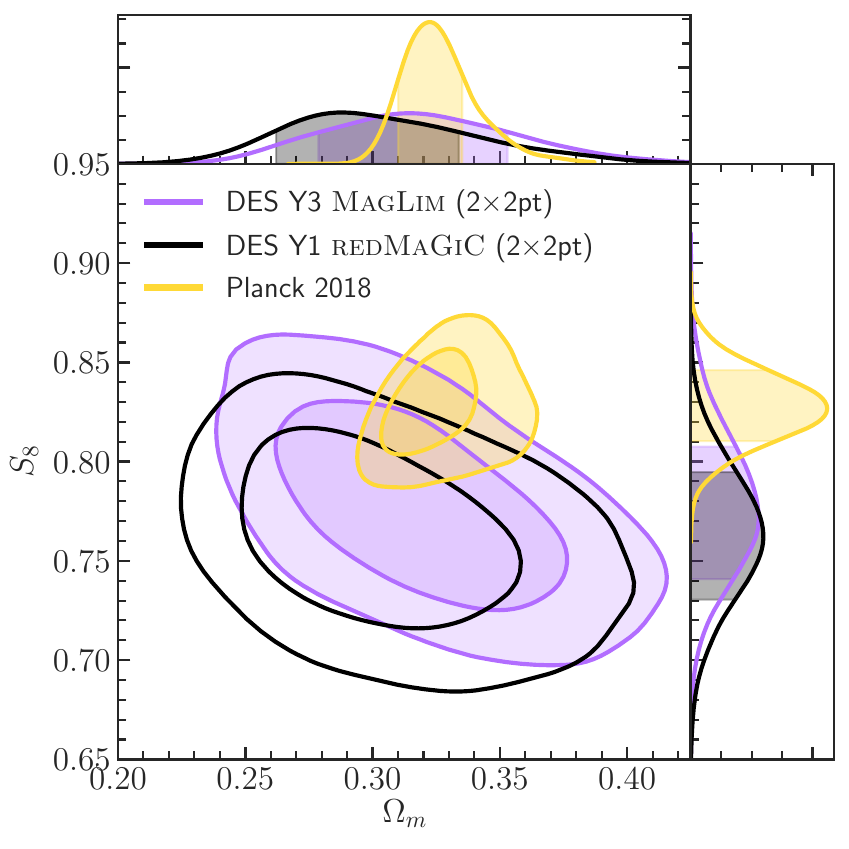}
		\caption{Constraints on $S_8$ and $\Omega_m$ in $\Lambda$CDM from galaxy-galaxy lensing and galaxy clustering using the \maglim\ sample. Also shown are the DES Y1 results and constraints from the \textit{Planck} CMB primary anisotropies.}
	\label{fig:lcdm_comp}
	\end{center}
\end{figure}

Our main results using galaxy clustering and galaxy-galaxy lensing for $\Lambda$CDM are shown in Fig.~\ref{fig:lcdm_comp} and Table~\ref{tab:post}, and compared with DES Y1 \cite{DESY1_3x2} and measurements of the CMB temperature and polarization anisotropies power spectra by the \textit{Planck} satellite \cite{2018arXiv180706209P}.
The \citet{y3-3x2ptkp} combines these two probes with cosmic shear to obtain our fiducial Y3 results. The constraints from Y3 are not significantly tighter than those from Y1, in spite of the factor of 3 gain in sky coverage. Much of this is due to several improvements in our modeling that were needed for the Y3 analysis due to the increased precision in our measurements, as described in \citet{y3-generalmethods}.

The marginalized 68\% C. L. mean values (best-fit values inside parentheses) of $S_8$, $\Omega_m$ and $\sigma_8$ are found to be
\begin{align}
S_8 &{} =  0.778^{+0.037}_{-0.031} \quad (0.809),  \\ 
\Omega_{\rm m} &{} =  0.320^{+0.041}_{-0.034} \quad (0.306), \\
\sigma_8 &{} =  0.758^{+0.074}_{-0.063} \quad (0.801) \, .
\end{align}

In the Appendix~\ref{sec: DV residuals}, we compare the theory prediction corresponding to these best-fit values with the measurements of galaxy clustering and galaxy-galaxy lensing for $\maglim$.

As mentioned before, in Fig.~\ref{fig:lcdm_comp} we compare the 2D marginalized constraints on $\Omega_m$ and $S_8$ to the \textit{Planck} CMB final release \cite{2018arXiv180706209P}. We include the primary temperature $TT$ data on scales $30 \le \ell \le 2508$, the E-mode and its cross power spectra with temperature ($EE+TE$) in the range $30 \le \ell \le 1996$, and the low-$\ell$ temperature and polarization likelihood ($TT +  EE$) at $2 \le \ell \le 29$. As done in the DES Y1 analysis \cite{DESY1_3x2}, we recompute the CMB likelihood in our fiducial parameter space from Table~\ref{tab:params}.

In order to quantify the level of agreement with \textit{Planck}, we calculate the Monte Carlo estimate of the probability of a parameter difference, presented in \cite{Raveri2020}. For this purpose, we compute the distribution of parameter shifts (in particular we consider $\Delta \sigma_8$ and $\Delta \Omega_m$) and estimate its compatibility with zero. We use the publicly available \texttt{tensiometer}\footnote{\url{https://github.com/mraveri/tensiometer}} code to compute the parameter shift between \textit{Planck} and our main DES-Y3 2$\times$2pt result in \LCDM, finding:
\begin{equation}
	\text{Parameter shift:} \, \, 1.0\sigma\,\, (\text{\maglim }\,\, 2\times2\text{pt vs \textit{Planck}} )
\end{equation}
which confirms more rigorously the qualitative agreement that we see in Fig.~\ref{fig:lcdm_comp}.

Even though the 2$\times$2pt constraints from DES Y1 are not independent, we can compute a rough estimate of the parameter shift with respect to  the 2$\times$2pt DES Y3 results by assuming no correlation between the two. We obtain a shift of $0.22\sigma$, confirming the good agreement that we see in Fig.~\ref{fig:lcdm_comp}.

\begin{figure}
	\begin{center}
		\includegraphics[width=\linewidth]{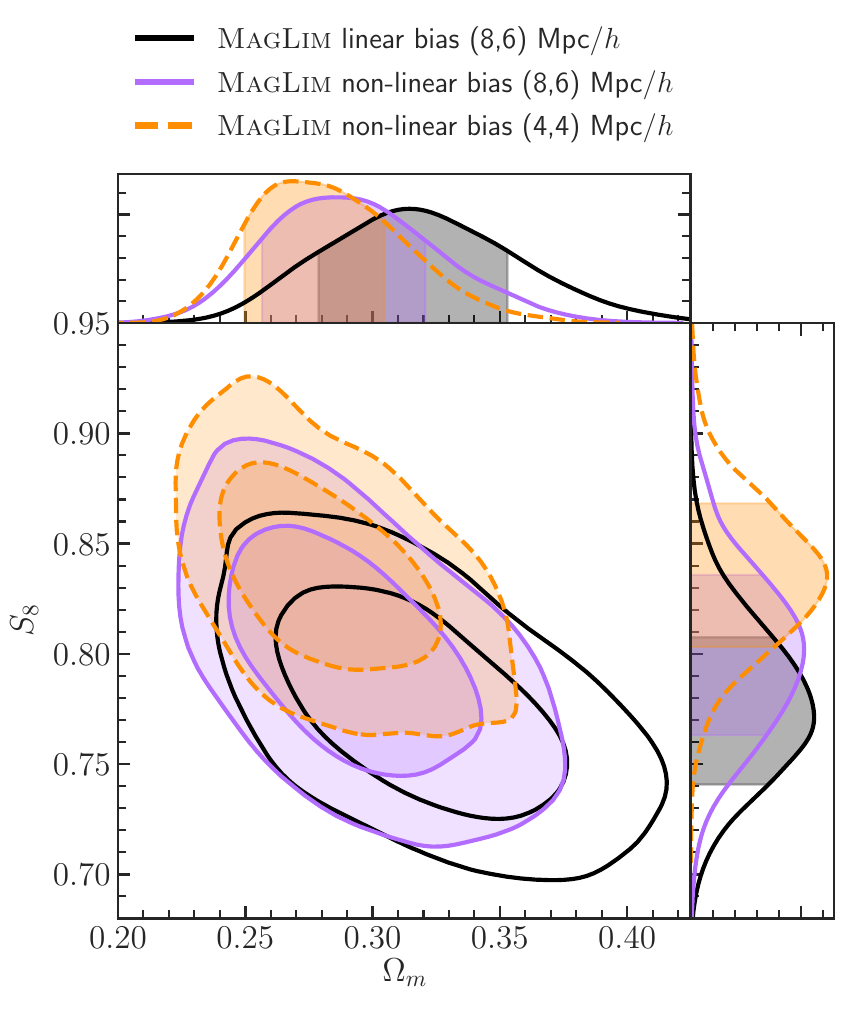}
		\caption{Same as Fig.~\ref{fig:lcdm_comp}, but the two new sets of contours use the non-linear bias model to analyze the data. This analysis agrees with the linear bias model when the same data vector is used $(8,6)$ but provides tighter constraints when smaller scales are included.}
		\label{fig: lcdm NL bias constraints}
	\end{center}
\end{figure}

Fig.~\ref{fig:sim_constraints} and related discussions in Sec.~\ref{sec:scale_cuts} show that including more information from small scales with a model that goes beyond linear bias recovers more cosmological information. We apply this non-linear bias model to the fiducial dataset that cuts scales below (8, 6) $h^{-1}$Mpc for ($w(\theta),\gamma_t$). The extended model applied to the same data as the linear bias model does not lose much constraining power as indicated by the purple contours in Fig.~\ref{fig: lcdm NL bias constraints} and the parameter values in Table~\ref{tab:post}. The contours in the $\Omega_m-S_8$ plane are shifted by  $0.4\sigma$ with respect to the linear bias analysis, therefore both galaxy bias models yield consistent results.

When opened up to include more small scale data (orange contours), up to (4, 4) $h^{-1}$Mpc, the analysis does provide tighter constraints. In particular, we obtain an improvement of $31\%$ in the $\Omega_{\rm m}-S_8$ plane. This is an indication that future analyses and surveys stand to benefit greatly from sophisticated modeling of small scales. The contours are shifted by $0.46\sigma$ with respect to the (8, 6) $h^{-1}$Mpc scale cuts using the same non-linear galaxy bias model, hence including smaller scales gives consistent results.

We have also tested the impact in the constraining power when including galaxy clustering cross-correlations in the analysis, finding a gain of 15\% in the $\Omega_{\rm m}-S_8$ plane. The clustering cross-correlations are much more sensitive to lens magnification than the auto-correlations, hence  \citet*{y3-2x2ptmagnification} further study the impact of magnification on the 2$\times$2pt cosmological constraints when including galaxy clustering cross-correlations.

\subsection{$w$CDM}

While \LCDM fits this 4-bin 2$\times$2pt dataset (and virtually all other datasets) well, in this section we consider  
its simplest possible extension, which is to allow its equation of state, $w\equiv P/\rho$, to differ from -1, that of the cosmological constant. This extension is dubbed $w$CDM.

\begin{figure}
	\begin{center}
		\includegraphics[width=\linewidth]{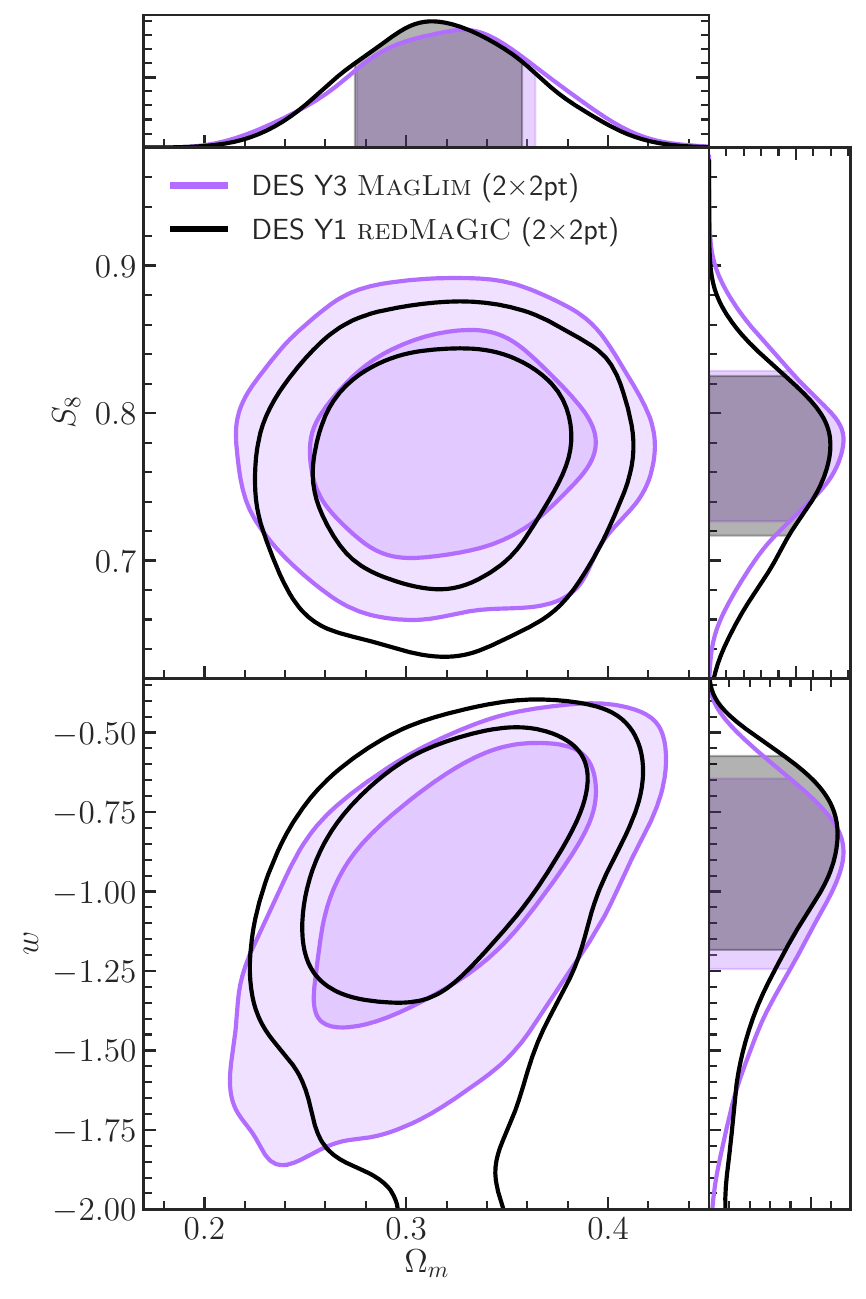}
		\caption{Constraints from the 2$\times$2pt data vector in $w$CDM. The range of $w$ on the vertical axis coincides with the prior on $w$, so not much information is added about $w$ from the 2$\times$2pt data.}
		\label{fig: wcdm constraint}
	\end{center}
\end{figure}

Fig.~\ref{fig: wcdm constraint} shows the constraints on $S_8, \Omega_m$ and $w$ in this model. The Y3 constraints are slightly tighter than those obtained from Y1, but by itself the 2$\times$2pt data are not very informative about the value of $w$. Recall that the prior imposed is $-2<w<-0.33$, so the information gained over the prior is modest.

\begin{figure}
	\begin{center}
		\includegraphics[width=\linewidth]{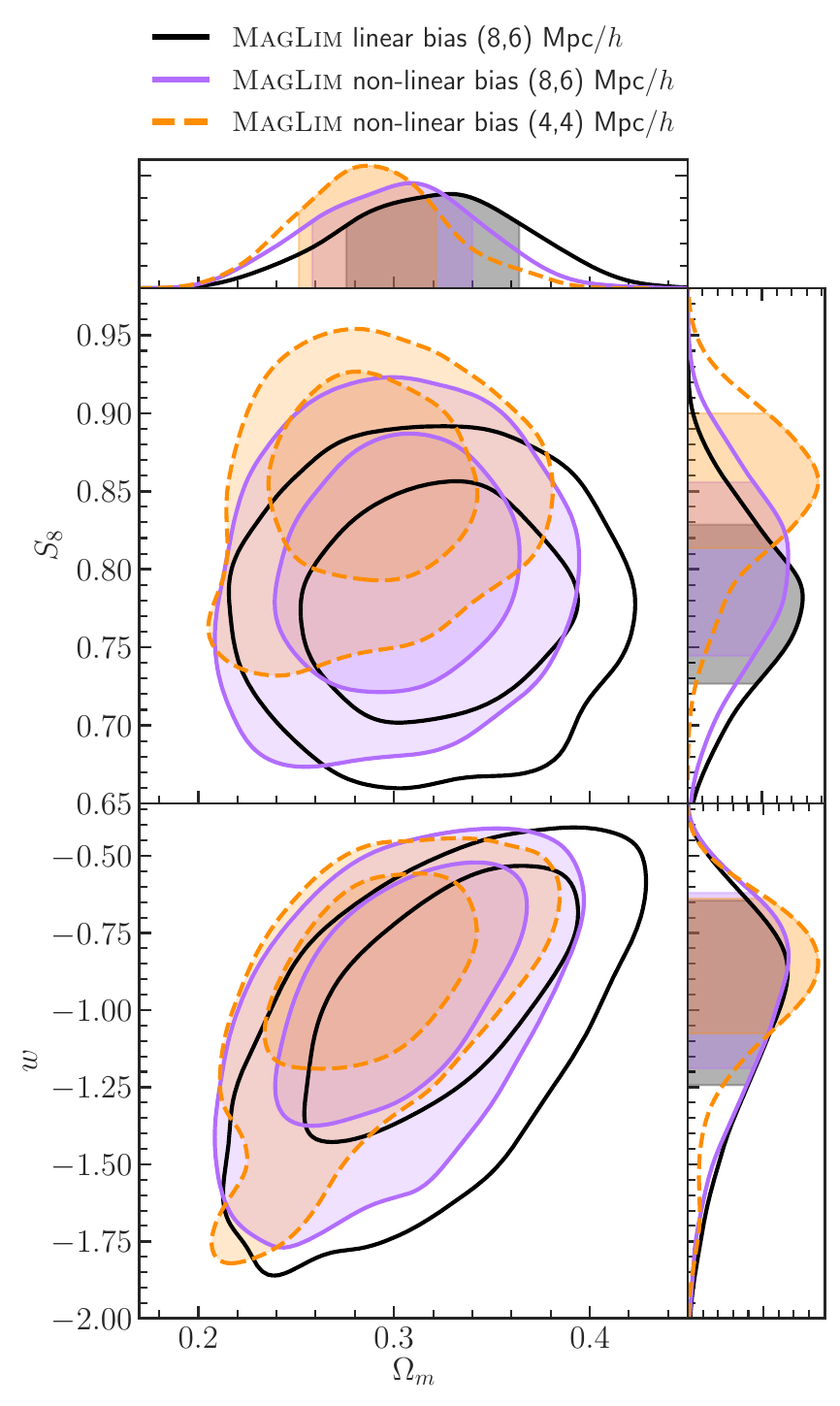}
		\caption{Same as Fig.~\ref{fig: wcdm constraint} but using smaller scales and the non-linear bias model.}
		\label{fig: wcdm NL bias constraints}
	\end{center}
\end{figure}

The results of the more aggressive analysis using smaller scales and the non-linear bias model are shown in Fig.~\ref{fig: wcdm NL bias constraints}. They are a bit tighter than the more conservative analysis (see Table~\ref{tab:post} for a quantitative comparison). In particular, we find improvements of 33\% in the $\Omega_m-S_8$ plane and about 41\% for $w-\Omega_m$.  
Despite the improvement on $w$, the lower panel in Fig.~\ref{fig: wcdm NL bias constraints} shows that the 2-sigma tails extend close to the prior boundary. Therefore,  2$\times$2pt by itself is not very constraining on the dark energy equation of state.

\begin{table*}
	\setlength{\extrarowheight}{7pt} 
	\caption{\label{tab:post} 68\% C.L. marginalized cosmological constraints in \LCDM and \wCDM  using the combination of DES Y3 galaxy clustering and galaxy-galaxy lensing measurements (2$\times$2pt).} 
	
	\begin{tabular*}{0.8\textwidth}{c @{\extracolsep{\fill}} lcccc}
		\hline
		\hline
		Cosmological model &  \ \ Galaxy bias model (scale cut) &$\Omega_m$ &$S_8$ &$\sigma_8$ & $w$ \\ 
		\hline
		\LCDM & \ \ linear bias (8,6) $\mpc$ & $0.320^{+0.041}_{-0.034}$ & $0.778^{+0.037}_{-0.031}$ & $0.758^{+0.074}_{-0.063}$ & ... \\ 
		\LCDM & \ \ non-linear bias (8,6) $\mpc$ & $0.293^{+0.037}_{-0.027}$ & $0.800^{+0.037}_{-0.036}$ & $0.813^{+0.074}_{-0.070}$ & ... \\ 
		\LCDM & \ \ non-linear bias (4,4) $\mpc$ & $0.284^{+0.034}_{-0.021}$ & $0.836^{+0.033}_{-0.033}$ & $0.863^{+0.071}_{-0.065}$ & ... \\ 
		\hline
		\wCDM & \ \ linear bias (8,6) $\mpc$ & $0.32^{+0.044}_{-0.046}$ & $0.777^{+0.049}_{-0.051}$ & $0.758^{+0.079}_{-0.061}$ & $-1.031^{+0.218}_{-0.379}$ \\ 
		\wCDM & \ \ non-linear bias (8,6) $\mpc$ & $0.301^{+0.043}_{-0.040}$ & $0.798^{+0.053}_{-0.058}$ & $0.802^{+0.076}_{-0.067}$ & $-0.993^{+0.197}_{-0.373}$ \\ 
		\wCDM & \ \ non-linear bias (4,4) $\mpc$ & $0.289^{+0.038}_{-0.033}$ & $0.849^{+0.036}_{-0.052}$ & $0.870^{+0.068}_{-0.067}$ & $-0.937^{+0.139}_{-0.299}$ \\ 
		\hline
		\hline
	\end{tabular*}
\end{table*}

\subsection{Robustness tests}
\label{sec:robust}

\begin{figure*}
	\begin{center}
		\includegraphics[width=0.9\linewidth]{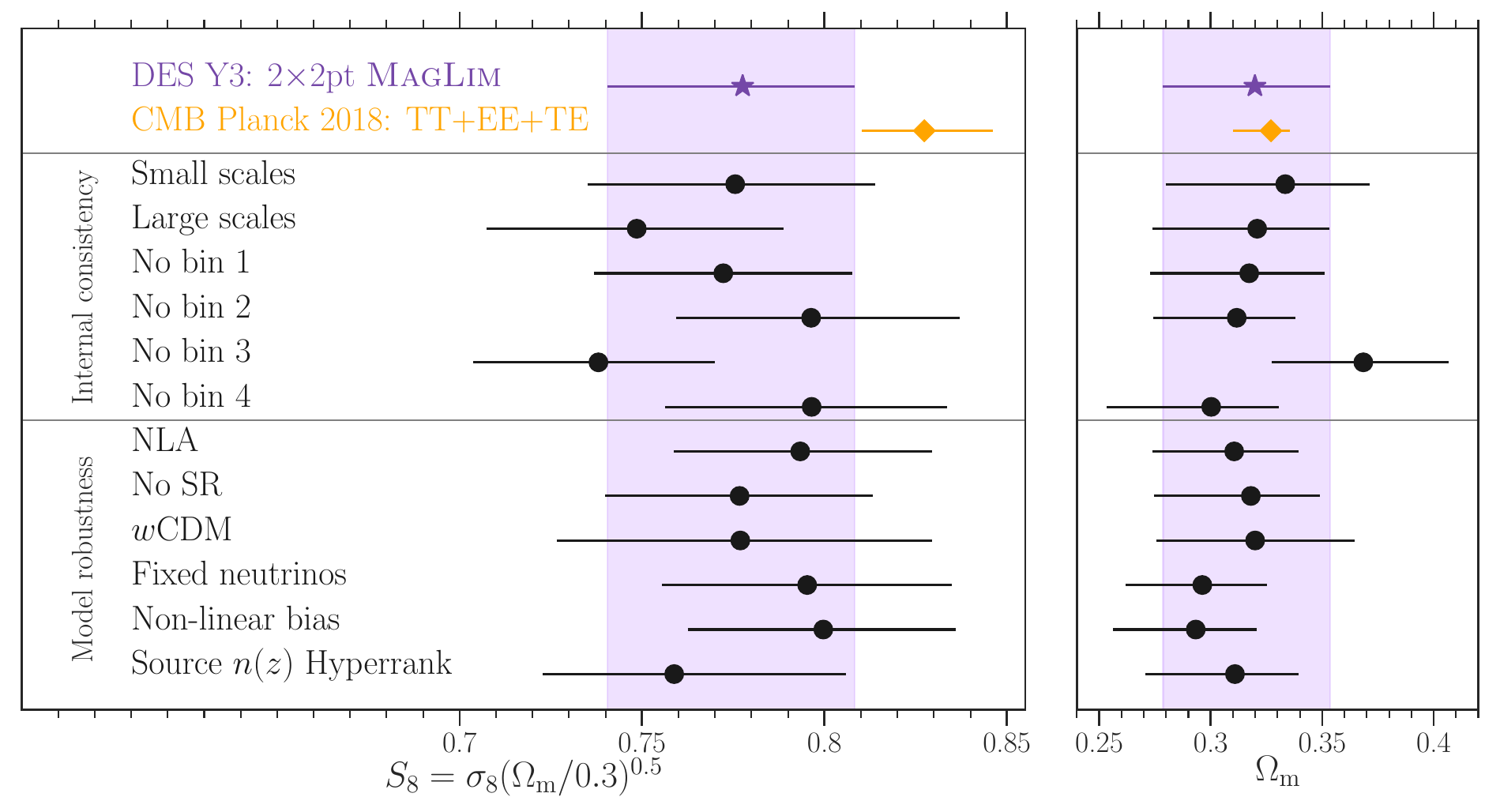}
		\caption{Comparison of the $\maglim$ 2$\times$2pt 68\% C.L. marginalized cosmological constraints in $\Lambda$CDM when changing the analysis choices.  }\label{fig: robustness tests}
	\end{center}
\end{figure*}

We assess the internal consistency of the data used in this analysis and robustness of the baseline model in $\Lambda$CDM in Fig.~\ref{fig: robustness tests}. The first row, as well as the shaded vertical bars, show the $1\sigma$ uncertainty on $S_8$ (left column) and $\Omega_m$ (right column) of the baseline $\Lambda$CDM  analysis presented in Sec.~\ref{sec:LCDM}. For reference, we also show the corresponding CMB constraints from \textit{Planck}, as described in Sec.~\ref{sec:LCDM} .

The next six rows, labeled \emph{internal consistency}, show the parameter constraints from different splits of the data vector: The \emph{small scales} and \emph{large scales} analyses restrict the analysis to angular scales corresponding to physical separations below/above $30\,\rm{Mpc}/h$; we note that the large-scale analysis does not marginalize over a point mass contribution of $\gamma_t$, as this term scales as $\theta^{-2}$ and thus insignificant for the large scales analysis. The next four rows show the parameter constraints when excluding one \maglim\ tomographic bin at a time. The largest parameter shift is caused by removing \maglim\ bin 3, but all data splits are consistent with the baseline result, pointing to the internal consistency of the data vector in $\Lambda$CDM. 

The bottom six rows, labeled \emph{model robustness}, show the parameter constraints for different analysis model variations:
\begin{itemize}
\item NLA: This analysis variation uses the non-linear alignment model for intrinsic alignments, which is a subspace of the baseline TATT model with $a_2 = \eta_2 = b_{\rm{TA}} = 0$.
\item No SR: This analysis variation does not include the shear ratio (SR) likelihood, which in the baseline analysis primarily adds constraining power on photometric redshift and intrinsic alignment parameters.
\item $w$CDM: This analysis variation shows the robustness of $S_8$ and $\Omega_\mathrm{m}$ constraints to the dark energy parameterization.
\item Fixed neutrino mass: This analysis variation fixes the neutrino mass to the minimum mass. As the neutrino mass is unconstrained (within prior range) by the baseline analysis, this variation primarily corresponds to a reduction in prior volume effects.
\item Non-linear bias: This analysis variation employs the non-linear bias model \emph{using the same scale cuts as the linear bias analysis}.
\item Source $n(z)$ Hyperrank: Here we account for the full-shape uncertainty in the source redshift distributions instead of marginalizing over an additive shift to the mean redshift. The method consists of sampling a large set of $n(z)$ realizations in the likelihood analysis with \textsc{Hyperrank} \cite{y3-hyperrank}.
\end{itemize}
The parameter constraints are consistent for all of these model variations, demonstrating the robustness of the baseline analysis choices.

Aside from the tests presented above, \citet{y3-galaxyclustering} shows that the constraints are robust to the method used to estimate the weights that correct our data vector from observational systematics. Furthermore, not correcting for the existing correlations with the survey property maps biases our results from galaxy clustering, which demonstrates the importance of estimating the weights accurately. See \cite{y3-galaxyclustering} for more details.

\section{Conclusions}
\label{sec:conclusions}

We have presented the DES-Y3 cosmological constraints obtained from the combination of galaxy clustering and galaxy-galaxy lensing (2$\times2$pt) using the $\maglim$ lens sample. The definition of this sample was previously optimized in \cite{y3-2x2maglimforecast} in terms of its forecasted $w$CDM cosmological constraints, with the goal of exploring the trade-off between number density and photometric redshift accuracy. It has 10.7 million galaxies comprising a redshift range between $z_{\min}=0.2$ and $z_{\max}=1.05$, which we split into 6 tomographic bins (see Table~\ref{tab:samples}). We use as sources the \metacal catalog \cite{y3-shapecatalog}, which consists of more than 100 million shapes divided in 4 tomographic bins (see Fig.~\ref{fig: lens source nzs}).

After validation of our modeling pipeline using both simulated theory data vectors and measurements from N-body simulations, we obtain our fiducial cosmological constraints using the first 4 tomographic bins of $\maglim$ (see Sec.~\ref{sec:cosmology-constrains} for further details). In \LCDM, we measure at 68\% C.L the clustering amplitude  $S_8 = 0.778^{+0.037}_{-0.031}$ and the matter energy density $\Omega_{\mathrm m}=0.320^{+0.041}_{-0.034}$. In $w$CDM, we obtain $S_8 = 0.777^{+0.049}_{-0.051}$,  $\Omega_{\mathrm m}=0.320^{+0.044}_{-0.046}$, and also constrain the dark energy equation of state $w=-1.031^{+0.218}_{-0.379}$.

We also extend our analysis to smaller scales by using a non-linear galaxy bias model, finding improvements of $31\%$ in the  $\Omega_{\mathrm m}-S_8$ plane in \LCDM and of $41\%$ for $w-\Omega_{\mathrm{m}}$ in $w$CDM. 

In Figs.~\ref{fig:lcdm_comp} and \ref{fig: wcdm constraint} we compare our fiducial results with DES Y1 2$\times$2pt, finding a very good agreement. In addition, we estimate the consistency of our \LCDM cosmological constraints with the results from the CMB from the \textit{Planck} satellite \cite[][TT+TE+EE]{2018arXiv180706209P}. We find that our constraints in the $\Omega_{\mathrm m}-S_8$ plane are low with respect to \textit{Planck} at the $1\sigma$ level. This result is in line with the slightly low $S_8$ values with respect to CMB  that have been measured in other weak lensing surveys, such as KiDS \cite{Joudaki2018,vanUitert2018} and HSC \cite{Hikage2019,Hamana2020}.

In Sec.~\ref{sec:robust} we evaluate the internal consistency of the $\maglim$ 2$\times$2pt results. We find that our 68\% C.L. constraints on $S_8$ and $\Omega_{\mathrm m}$ are consistent across angular scales, tomographic bins, and modeling analysis choices. The results presented here are complemented by the equivalent 2$\times$2pt analysis using the 
$\redmagic$ sample \cite{y3-2x2ptbiasmodelling}, a study of the impact of magnification on the 2$\times$2pt cosmological constraints \cite{y3-2x2ptmagnification},  and the cosmic shear analysis in \cite{y3-cosmicshear1,y3-cosmicshear2}. The $\maglim$ 2$\times$2pt measurements are combined with cosmic shear in \cite{y3-3x2ptkp} to obtain the DES-Y3 fiducial cosmological results.

The advances in methodology implemented in DES Y3 set the stage for the analysis of the full DES dataset, comprising 6 years of observations. Regarding the lens samples, the improvements include the optimization of the sample in terms of its cosmological constraints \cite{y3-2x2maglimforecast}, a full characterization of the uncertainties in the redshift distributions using self-organizing maps \cite{y3-2x2ptaltlenssompz}, and the inclusion of several upgrades to our modeling \cite{y3-generalmethods}, such as  lens magnification \cite{y3-2x2ptmagnification}, point-mass marginalization \cite{y3-2x2ptbiasmodelling}, and non-linear galaxy bias \cite{y3-2x2ptbiasmodelling}. These advances  will be critical for the future `Stage IV' photometric surveys such as Euclid \cite{Euclid}, the Nancy G. Roman Space Telescope \cite{Spergel2015}, and the Vera C. Rubin Observatory Legacy Survey of Space and Time (LSST) \cite{LSST}.

\input{acknowledgements.tex}

\bibliographystyle{apsrev4-2}
\bibliography{des_y3kp.bib,maglim.bib}

 \appendix
 
  \section{Comparison of DNF photometric redshifts with matched spectroscopic data}
 \label{sec: Comparison of DNF with VIPERS}

  \begin{figure}[ht]
	\begin{center}
		\includegraphics[width=\linewidth]{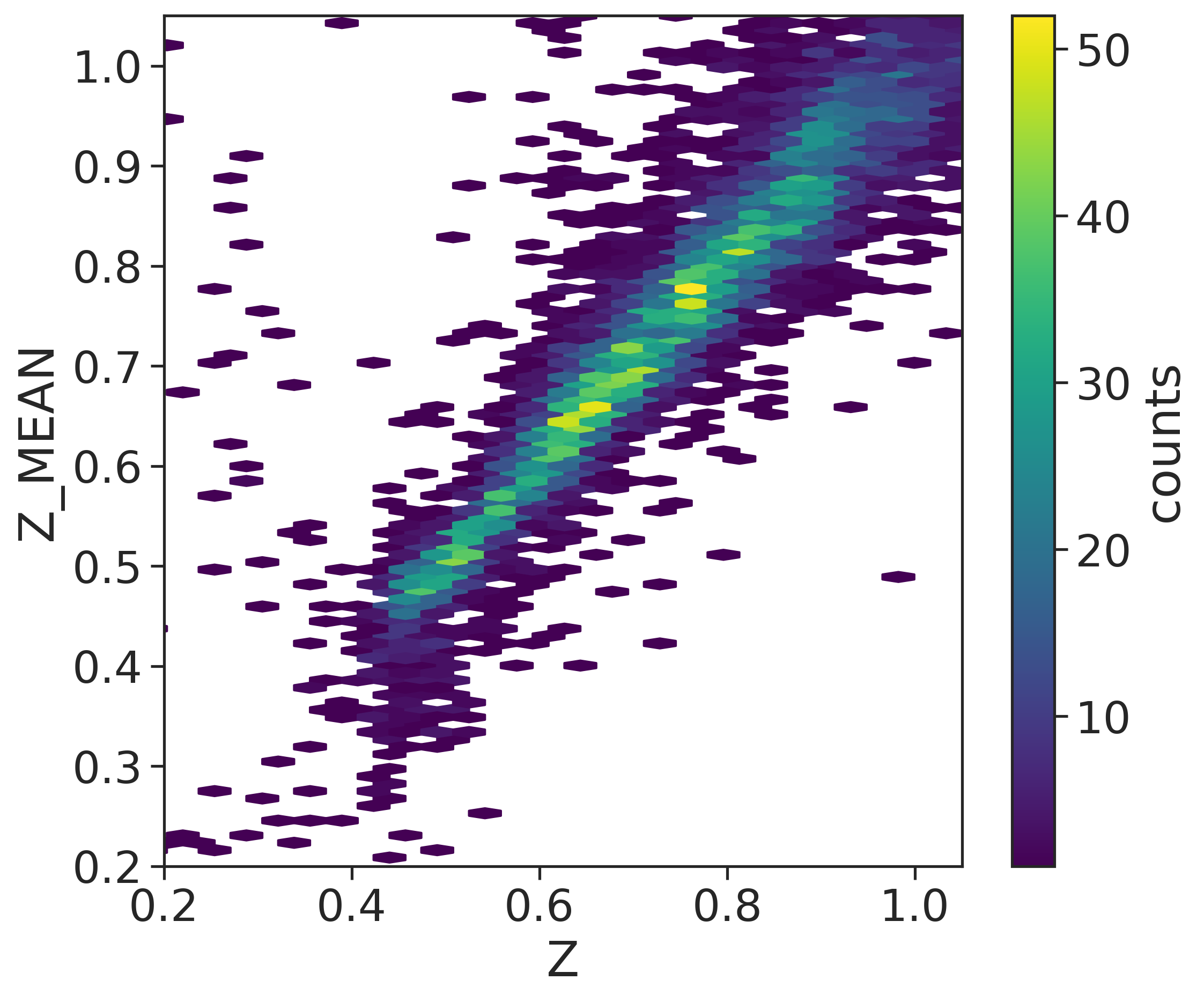}
		\caption{\maglim DNF photometric redshift estimates (Z\_MEAN) compared to the spectroscopic redshifts from VIPERS (Z).}
		\label{fig: photoz vs specz}
	\end{center}
\end{figure}

As explained in Sec.~\ref{sec: dnf}, we use \dnf\, to select the $\maglim$ galaxies and assign them to tomographic bins. \dnf\, computes a point estimate of the
true redshift by performing a fit to a hyperplane using 80 nearest neighbors in color and
magnitude space taken from a reference set that has an
associated true redshift from a large spectroscopic database. In Fig.~\ref{fig: photoz vs specz}, we compare the \dnf\, photometric point estimates (Z\_MEAN) to the spectroscopic redshifts from VIPERS. We use VIPERS for this validation because it was not included in the training set for the \dnf\, estimation of the point estimates. However, VIPERS is only complete at $z > 0.5$. Therefore, the comparison is limited by the low number of matched galaxies at lower redshifts. Using VIPERS as reference, we estimate the following two and three $\sigma$ outlier rates:
\begin{equation}
\begin{gathered}
	2\sigma \, \mathrm{outlier\, rate} = 0.128, \\
	3\sigma\,  \mathrm{outlier\,  rate} = 0.073,
\end{gathered}
\end{equation}
where the outliers in the $2\sigma$ region, for instance, are selected with
\begin{equation}
| \mathrm{Z\_MEAN} - \mathrm{Z}| > 2\sigma_{68}(\mathrm{Z\_MEAN} - \mathrm{Z}),
\end{equation}
and $\sigma_{68}(\mathrm{Z\_MEAN} - \mathrm{Z})$ is the 68\% confidence interval of values in the distribution of $(\mathrm{Z\_MEAN} - \mathrm{Z})$.
 
 Since the \maglim selection includes a fraction of blue galaxies, we make a color split in order to compare the photometric quality of the blue galaxies compared to the full \maglim sample. For simplicity, we select the blue galaxies from the MagLim sample by applying the inverse of the color cut used for the DES BAO sample, which is optimized to select the reddest galaxies \cite{DESY1BAOsample}. Therefore, we apply this color cut to the MagLim sample that depends on the $i$, $z$ and $r$ band magnitudes: $(i - z) +2.0(r - i) < 1.7 $. The resulting sample has 4,953,191 galaxies, about 46\% of the total MagLim sample. The outlier rates are very similar to the full sample. In order to estimate the redshift uncertainty, $\sigma_z/(1+z)$, we use the 68\% confidence interval of values in the distribution of $(\mathrm{Z\_MEAN} - \mathrm{Z})/(1 + \mathrm{Z})$ around its median value. For the full MagLim sample, we find $\sigma_z/(1+z) = 0.027$, while for the fraction of blue galaxies, we find $\sigma_z/(1+z) = 0.037$.
 
In order to validate the performance of $\dnf$ for the MagLim blue galaxies at lower redshifts ($z<0.5$), we use the spectroscopic data set from \cite{Gschwend:2018}. This reference catalog contains about $2.2\times 10^5$ spectra matched to DES objects from 24
 	different spectroscopic catalogs, including Sloan Digital
 	Sky Survey (SDSS) DR14 \cite{sdssdr14}, DES’s own
 	follow-up through the OzDES program \cite{ozdes}, and the Galaxy and Mass Assembly survey (GAMA) \cite{GAMA}. Using this data set, we find similar values for the outlier rates and  $\sigma_z/(1+z)$ compared to using VIPERS as a reference. However, we note that the outlier rate and the redshift uncertainty $\sigma_z/(1+z)$ are higher in the photometric redshift range $[0.5, 0.6]$. The $3\sigma$ outlier rate reaches significantly higher values in that interval, $3\sigma$ $\sim 0.12$, whereas $\sigma_z/(1+z)\sim 0.03$. This increase in the outliers is caused by degeneracies among galaxy types at $z\sim0.4$, which are difficult to break due to the lack of $u$ band information in DES. 
 	As explained in \cite{y3-gold}, this spectroscopic catalog has been used as training data for \dnf, and thus it is not an independent data set like VIPERS.

 \section{Results after unblinding}
 \label{sec: appendix unblinding results}
 
  \begin{figure}[ht]
 	\begin{center}
 		\includegraphics[width=\linewidth]{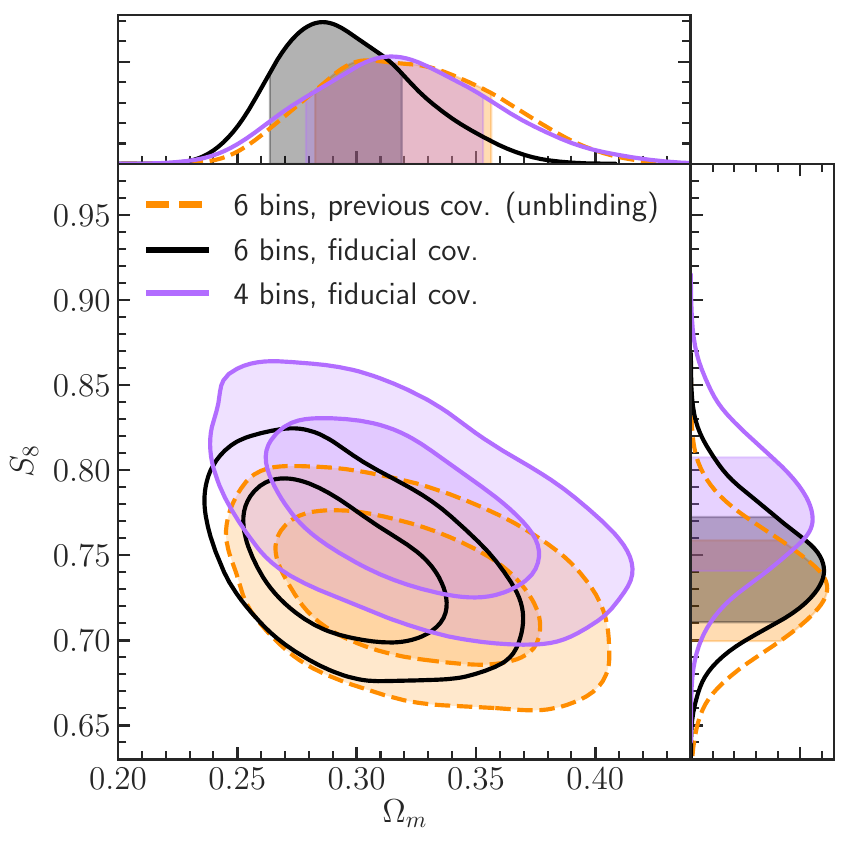}
 		\caption{Comparison of 2$\times$2pt $\Lambda$CDM constraints for $\maglim$ using different iterations of the covariance matrix and number of tomographic bins included. The dashed orange lines correspond to the unblinding results, in which we used the whole set of bins and a covariance that assumed some fiducial cosmology and galaxy bias. Solid black corresponds to the results after updating the covariance with the 3$\times$2pt $\Lambda$CDM best-fit cosmology. Last, the purple filled contours show the fiducial 2$\times$2pt constraints, with just the first four tomographic bins included in the analysis. }
 		\label{fig: cov comparison lcdm}
 	\end{center}
 \end{figure}

\begin{figure}
	\begin{center}
		\includegraphics[width=\linewidth]{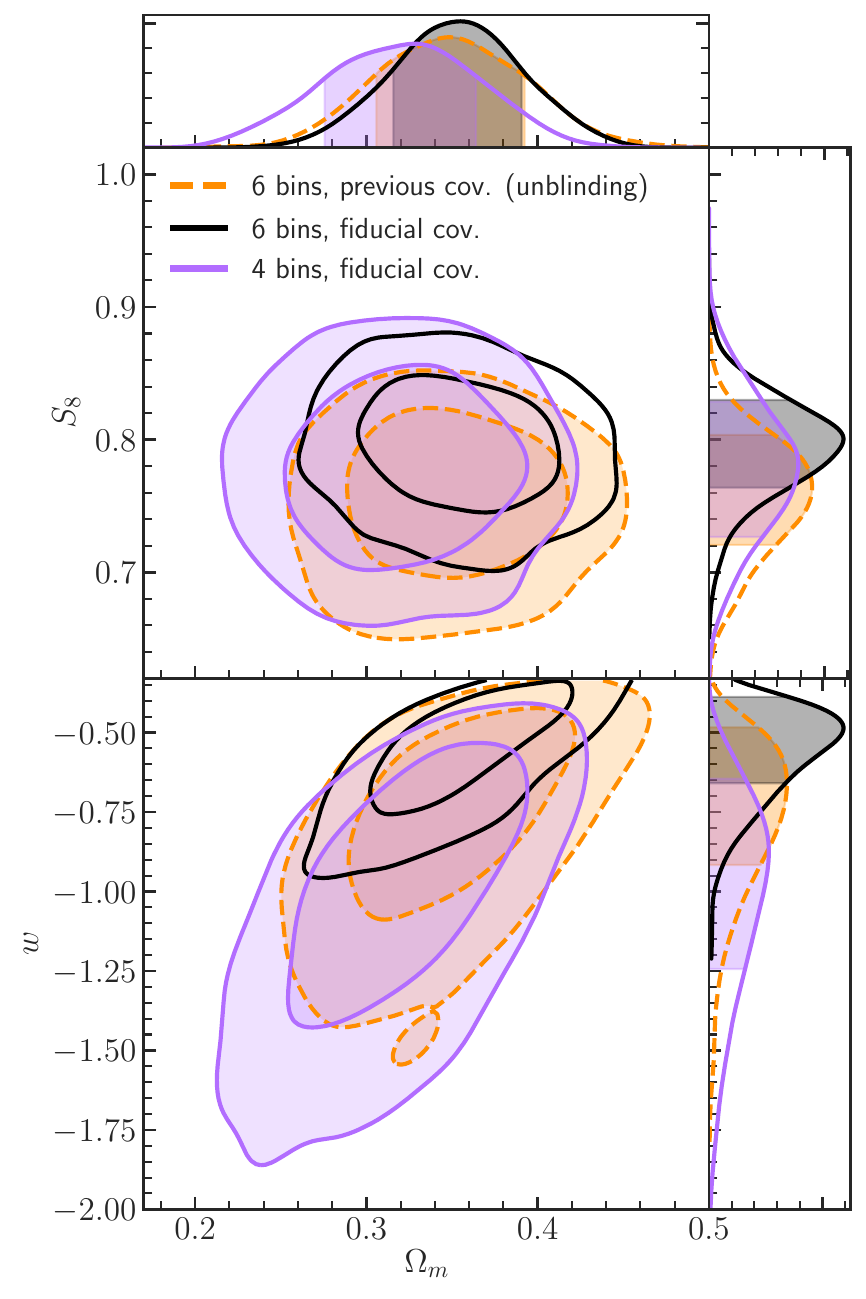}
		\caption{Comparison of 2$\times$2pt $w$CDM constraints for $\maglim$ using different iterations of the covariance matrix and number of tomographic bins included.  }
		\label{fig: cov comparison wcdm}
	\end{center}
	\label{fig:unblinding-states-wCDM}
\end{figure}

\begin{table*}
	\setlength{\extrarowheight}{7pt} 
	\caption{\label{tab: unblinding constraints} 68\% C.L. marginalized cosmological constraints in \LCDM and \wCDM  using different iterations of the covariance matrix and number of tomographic bins.} 
	
	\begin{tabular*}{0.85\textwidth}{c @{\extracolsep{\fill}} lcccc}
		\hline
		\hline
		Cosmological model &  \qquad \qquad \qquad Iteration &$\Omega_m$ &$S_8$ &$\sigma_8$ & $w$ \\ 
		\hline
		\LCDM & 6 bins, previous cov. (unblinding)  & $0.322^{+0.040}_{-0.034}$ & $0.729^{+0.030}_{-0.030}$ & $0.707^{+0.063}_{-0.053}$ & ... \\ 
		\LCDM & 6 bins, fiducial cov. & $0.297^{+0.033}_{-0.022}$ & $0.744^{+0.034}_{-0.029}$ & $0.751^{+0.061}_{-0.056}$ & ... \\ 
		\LCDM & 4 bins, fiducial cov.   & $0.320^{+0.041}_{-0.034}$ & $0.778^{+0.037}_{-0.031}$ & $0.758^{+0.074}_{-0.063}$ & ... \\ 
		\hline
		\wCDM & 6 bins, previous cov. (unblinding)  & $0.349^{+0.044}_{-0.043}$ & $0.755^{+0.035}_{-0.049}$ & $0.703^{+0.064}_{-0.053}$ & $-0.789^{+0.128}_{-0.306}$ \\ 
		\wCDM & 6 bins, fiducial cov.   & $0.353^{+0.037}_{-0.038}$ & $0.794^{+0.030}_{-0.037}$ & $0.735^{+0.059}_{-0.049}$ & $-0.58^{+0.081}_{-0.193}$ \\ 
		\wCDM & 4 bins, fiducial cov.  & $0.320^{+0.044}_{-0.046}$ & $0.777^{+0.049}_{-0.051}$ & $0.758^{+0.079}_{-0.061}$ & $-1.031^{+0.218}_{-0.379}$ \\ 
		\hline
		\hline
	\end{tabular*}
\end{table*}

As discussed in Sec.\ref{sec:cosmology-constrains} \maglim passed all the unblinding requirements and produced at first good model fits to both \LCDM and \wCDM. The results from these fits however indicated a considerably lower linear galaxy bias, in particular for the last tomographic bins. Updating the covariance with these values resulted in a much tighter errors, which in turned implied that the cosmological chains with the updated covariance violated the $\chi^2$ thresholds in \LCDM and \wCDM. 

Different tests showed that the model had particular trouble in providing a consistent fit to both clustering and galaxy-galaxy lensing amplitudes on the last two tomographic bins (i.e. from 0.85 to 1.05). This can be visualized by comparing the $\sigma_8$ and galaxy bias values obtained from the combination of cosmic shear and galaxy-galaxy lensing ($\xi_\pm + \gamma_t$), cosmic shear and galaxy clustering ($\xi_\pm + w(\theta)$), or the whole  three two-point functions ($3\times2$pt, $\xi_\pm + \gamma_t+w(\theta)$) from \cite{y3-3x2ptkp},
\begin{itemize}
\item{$\xi_\pm + \gamma_t$: \\  $\sigma_8 = 0.61, b^i =\{ 1.81, 2.0, 2.48, 2.07, 1.49, 1.05\}$},
\item{$\xi_\pm + w$: \\ $\sigma_8 =  0.86, b^i = \{ 1.36,1.59, 1.63, 1.56, 1.71, 1.68\}$},
\item{{\rm $3\times2$pt} : \\ $\sigma_8 = 0.733, b^i = \{1.42,1.66,1.92,1.78,1.97,1.74\}$}.
\end{itemize}
 In terms of clustering amplitudes $b^i \sigma_8$ with respect to the corresponding best-fit values from $3\times2$pt in \LCDM we obtain, 
\begin{itemize}
\item{$\xi_\pm + w$ / $3\times2$pt : \\ $b^i \sigma_8 = \{ 1.12, 1.12, 1.00, 1.02, 1.01, 1.13\}$},  
\item{$\xi_\pm + \gamma_t$ / $3\times2$pt: \\ $b^i \sigma_8 = \{1.06, 1.00, 1.07, 0.96, 0.63, 0.5\}$}.
\end{itemize}
This seems to indicate that the galaxy-galaxy lensing signal is too low with respect to the theory model in the last two bins, leading to unexpectedly low values for the galaxy biases
given the brightness of the \maglim sample as well as low $\sigma_8$. 
We have investigated and discarded various potential sources for this discordance, such as observational systematics \cite{y3-galaxyclustering} or underestimations of the width parameters in the redshift distributions. We have also discarded any correlation of this bias discordance with a particular region of the footprint \cite{y3-2x2ptbiasmodelling}. The last bins are the ones most impacted by magnification \cite{y3-2x2ptmagnification}, and an overestimation of the magnification coefficients (predicted magnification being too high) could indeed be driving at least part of this effect. In Fig. 7 of \cite{y3-gglensing}, the magnification contribution at these bins is already similar to the full galaxy-galaxy lensing signal. In the Appendix~\ref{sec: DV residuals}, we show the residuals between the 2$\times$2pt measurements and the best-fit theory model when using all 6 tomographic bins. The last 2 bins show some fluctuations in the measurements that contribute to the poor fit of the model ($p<10^{-3}$).  In all, these bins do not carry much signal-to-noise for the $3\times2$pt combination, and we decided to simply remove them from the analysis \cite{y3-3x2ptkp}. We defer further investigations to subsequent work.

The summary of best-fit $\chi^2$ values and goodness-of-fit PPD $p$-values for these three sets of chains following the initial unblinding are:
\vspace{3pt}
 \begin{itemize} 
 	\item 2$\times$2pt LCDM unblinding, 6 bins: $\chi^2=367$ for 336 degrees of freedom\footnote{We have not accounted for the informative priors used on the parameters when estimating the number of degrees of freedom. We expect the effective number of degrees of freedom to be lower. For example, for the cosmic shear analysis \cite{y3-cosmicshear1,y3-cosmicshear2} the effective number of free parameters is $\sim$5, compared to the total number of free parameters (28).} (373 data points and 37 free parameters), calibrated PPD  $p=0.351$,
 	\item 2$\times$2pt LCDM fiducial cov., 6 bins: $\chi^2=463$ for 336 degrees of freedom (373 data points and 37 free parameters), calibrated PPD  $p<10^{-3}$,
 	\item 2$\times$2pt LCDM fiducial cov., 4 bins: $\chi^2=280$ for 204 degrees of freedom (235 data points and 31 free parameters), calibrated PPD  $p=0.019$,
 	\vspace{-2pt}
\end{itemize}
where the first entry corresponds to the unblinding result, the second to those using a covariance updated to the unblinding best-fit results (and keeping the full data-vector), and the third to a re-run removing the last two lens bins. A further update to the covariances does not change the best fit values or the goodness-of-fit, hence the updated covariance after unblinding became our fiducial covariance. As discussed, the first and third stages satisfy the pre-established PPD threshold for publication of $p > 0.01$. 

Furthermore, in Table~\ref{tab: unblinding constraints} we detail the mean and $68\%$ C.L. for the relevant cosmological parameters, at each stage, for both \LCDM and \wCDM. Removing the last two bins decreases the constraining power considerably, by $27\%$ ($38\%$) in the $\Omega_m-S_8$ plane in \LCDM (\wCDM) and by 62\% in the $\Omega_m-w$ plane. However, the changes in the mean of the parameters are within the corresponding final $68\%$ C.L. region (i.e. within 1$\sigma$) when removing the last two bins. Except for $w$, that moves by 1.5$\sigma$'s towards \LCDM. In all, we conclude that the cosmology results are not driven by the decision on removing the last two tomographic bins from the data vector. In Figs. \ref{fig: cov comparison lcdm} and \ref{fig: cov comparison wcdm} we show the two dimensional contour plots for the posteriors in ($S_8,\Omega_m$) for \LCDM and ($S_8,\Omega_m,w$) for \wCDM, respectively.

\vspace{-0.5cm}

\section{Data-vector residuals}
\label{sec: DV residuals}
\begin{figure*}[ht]
	\begin{center}
		\includegraphics[width=\linewidth]{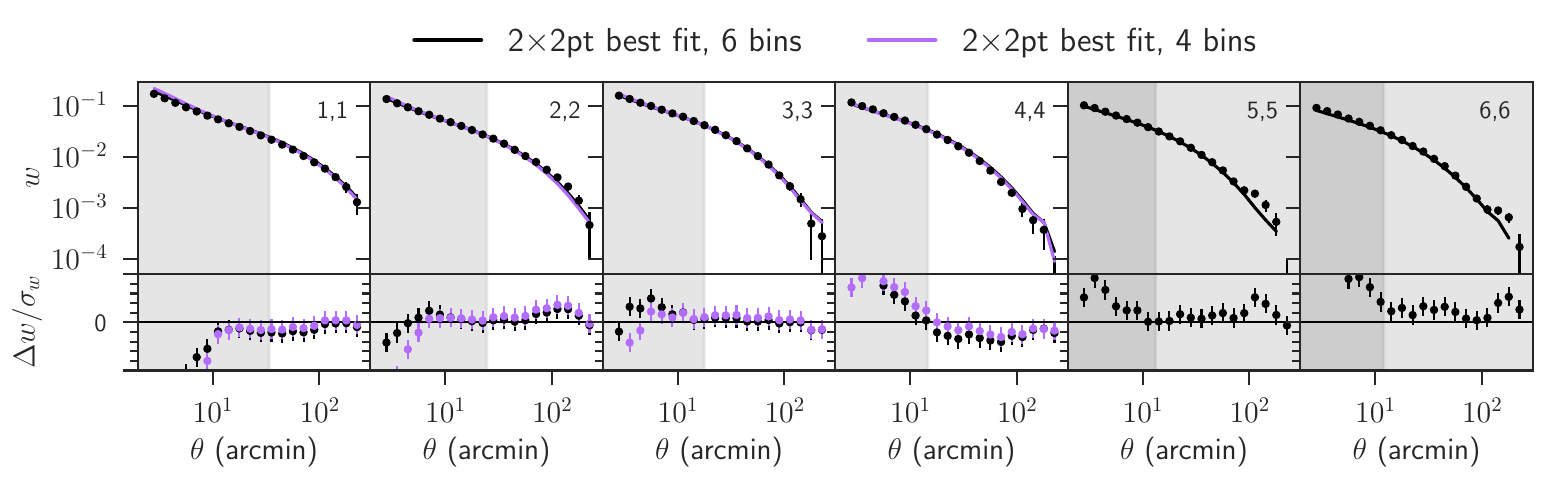}
		\caption{Galaxy clustering $w(\theta)$ measurements for $\maglim$ in each tomographic bin $i$. The best-fit \LCDM model from the 2$\times$2pt analysis using all 6 redshift bins is shown with a solid black line, while the best-fit when using the first 4 bins (the fiducial result) is shown in purple. The bottom part of each panel shows the fractional difference between the measurements and the model prediction, $(w^{\rm obs.}-w^{\rm th.})/\sigma_w$, with the $y$-axis range being $\pm5\sigma$. The angular scales excluded in the analysis are shaded, and bins 5 and 6 are not included in the final analysis. }
		\label{fig: wtheta DV}
	\end{center}
\end{figure*}

\begin{figure*}
	\begin{center}
		\includegraphics[width=\linewidth]{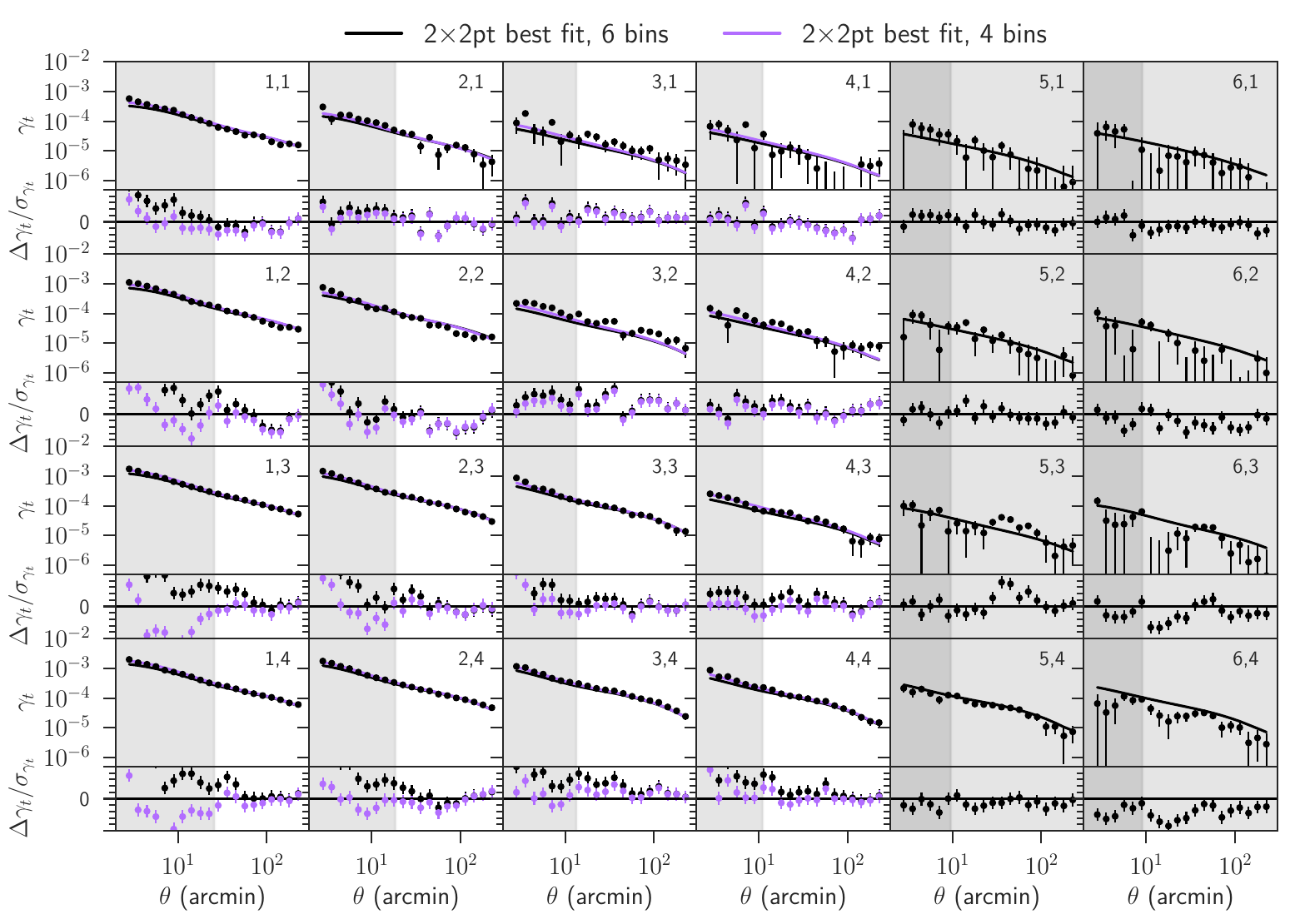}
		\caption{ Same as Fig.~\ref{fig: wtheta DV} but showing the measured tangential shear $\gamma_t(\theta)$. In each panel, the label $i, j$ refers to the lens tomograhic bin $i$ and the source bin $j$. }
		\label{fig: gammat DV}
	\end{center}
\end{figure*}

\vspace{-0.2cm}
We show the measurements for galaxy clustering $w(\theta)$ and tangential shear $\gamma_t(\theta)$ using the $\maglim$ lens sample. These measurements are described in detail in \citet{y3-galaxyclustering} and \citet{y3-gglensing}, respectively, and summarized in Sec.~\ref{subsec: DVs}. In Figs.~\ref{fig: wtheta DV} and \ref{fig: gammat DV}, we compare these measurements with the best-fit \LCDM theory prediction when using all 6 tomographic bins (black lines) or just the first 4 bins (purple), which is our final result. 

\end{document}

%% file: acknowledgements.tex
\begin{acknowledgements}

Funding for the DES Projects has been provided by the U.S. Department of Energy, the U.S. National Science Foundation, the Ministry of Science and Education of Spain, 
the Science and Technology Facilities Council of the United Kingdom, the Higher Education Funding Council for England, the National Center for Supercomputing 
Applications at the University of Illinois at Urbana-Champaign, the Kavli Institute of Cosmological Physics at the University of Chicago, 
the Center for Cosmology and Astro-Particle Physics at the Ohio State University,
the Mitchell Institute for Fundamental Physics and Astronomy at Texas A\&M University, Financiadora de Estudos e Projetos, 
Funda{\c c}{\~a}o Carlos Chagas Filho de Amparo {\`a} Pesquisa do Estado do Rio de Janeiro, Conselho Nacional de Desenvolvimento Cient{\'i}fico e Tecnol{\'o}gico and 
the Minist{\'e}rio da Ci{\^e}ncia, Tecnologia e Inova{\c c}{\~a}o, the Deutsche Forschungsgemeinschaft and the Collaborating Institutions in the Dark Energy Survey. 

The Collaborating Institutions are Argonne National Laboratory, the University of California at Santa Cruz, the University of Cambridge, Centro de Investigaciones Energ{\'e}ticas, 
Medioambientales y Tecnol{\'o}gicas-Madrid, the University of Chicago, University College London, the DES-Brazil Consortium, the University of Edinburgh, 
the Eidgen{\"o}ssische Technische Hochschule (ETH) Z{\"u}rich, 
Fermi National Accelerator Laboratory, the University of Illinois at Urbana-Champaign, the Institut de Ci{\`e}ncies de l'Espai (IEEC/CSIC), 
the Institut de F{\'i}sica d'Altes Energies, Lawrence Berkeley National Laboratory, the Ludwig-Maximilians Universit{\"a}t M{\"u}nchen and the associated Excellence Cluster Universe, 
the University of Michigan, the National Optical Astronomy Observatory, the University of Nottingham, The Ohio State University, the University of Pennsylvania, the University of Portsmouth, 
SLAC National Accelerator Laboratory, Stanford University, the University of Sussex, Texas A\&M University, and the OzDES Membership Consortium.

Based in part on observations at Cerro Tololo Inter-American Observatory, National Optical Astronomy Observatory, which is operated by the Association of 
Universities for Research in Astronomy (AURA) under a cooperative agreement with the National Science Foundation.

The DES data management system is supported by the National Science Foundation under Grant Numbers AST-1138766 and AST-1536171.
The DES participants from Spanish institutions are partially supported by MINECO under grants AYA2015-71825, ESP2015-66861, FPA2015-68048, SEV-2016-0588, SEV-2016-0597, and MDM-2015-0509, 
some of which include ERDF funds from the European Union. IFAE is partially funded by the CERCA program of the Generalitat de Catalunya.
Research leading to these results has received funding from the European Research
Council under the European Union's Seventh Framework Program (FP7/2007-2013) including ERC grant agreements 240672, 291329, and 306478.
We  acknowledge support from the Australian Research Council Centre of Excellence for All-sky Astrophysics (CAASTRO), through project number CE110001020, and the Brazilian Instituto Nacional de Ci\^encia
e Tecnologia (INCT) e-Universe (CNPq grant 465376/2014-2).

This manuscript has been authored by Fermi Research Alliance, LLC under Contract No. DE-AC02-07CH11359 with the U.S. Department of Energy, Office of Science, Office of High Energy Physics. The United States Government retains and the publisher, by accepting the article for publication, acknowledges that the United States Government retains a non-exclusive, paid-up, irrevocable, world-wide license to publish or reproduce the published form of this manuscript, or allow others to do so, for United States Government purposes.
Computations were made on the supercomputer Guillimin from McGill University, managed by Calcul Qu\'{e}bec and Compute Canada. The operation of this supercomputer is funded by the Canada Foundation for Innovation (CFI), the minist\`{e}re de l'\'{E}conomie, de la science et de l'innovation du Qu\'{e}bec (MESI) and the Fonds de recherche du Qu\'{e}bec - Nature et technologies (FRQ-NT).
This research is part of the Blue Waters sustained-petascale computing project, which is supported by the National Science Foundation (awards OCI-0725070 and ACI-1238993) and the state of Illinois. Blue Waters is a joint effort of the University of Illinois at Urbana-Champaign and its National Center for Supercomputing Applications.

This research used resources of the Ohio Supercomputer
Center (OSC) \cite{OSC} and of the National Energy Research
Scientific Computing Center (NERSC), a U.S. Department of
Energy Office of Science User Facility operated under Contract
No. DE-AC02-05CH11231.

\end{acknowledgements}